\documentclass[fleqn,usenatbib]{mnras}

\usepackage{newtxtext,newtxmath}

\usepackage[T1]{fontenc}

\DeclareRobustCommand{\VAN}[3]{#2}
\let\VANthebibliography\thebibliography
\def\thebibliography{\DeclareRobustCommand{\VAN}[3]{##3}\VANthebibliography}

\usepackage{graphicx}	
\usepackage{amsmath}	
\usepackage{tabularx}
\usepackage{xcolor}

\newcolumntype{Y}{>{\centering\arraybackslash}X}

\title[Dusty Tori in High $L/L_\mathrm{Edd}$ AGNs]{I Zw 1 and H0557-385: The Dusty Tori of Two High Eddington AGNs Observed in the MATISSE $LM$-Bands}

\author[F. Drewes et al.]{
Farin Drewes$^{1}$\thanks{E-mail: \href{mailto:n.c.drewes@soton.ac.uk}{n.c.drewes@soton.ac.uk}}, James H. Leftley$^{2}$, Sebastian F. H\"{o}nig$^{1}$, Konrad R. W. Tristram$^{3}$, Makoto Kishimoto$^{4}$
\\
$^{1}$School of Physics \& Astronomy, University of Southampton, University Road, Southampton SO17 1BJ, UK\\
$^{2}$Université Côte d’Azur, Observatoire de la Côte d’Azur, CNRS, Laboratoire Lagrange, Boulevard de l’Observatoire, CS 34229,
06304 Nice Cedex 4, France\\
$^{3}$European Southern Observatory, Alonso de Córdova 3107, Casilla 19001, Santiago, Chile\\
$^{4}$Department of Astrophysics \& Atmospheric Sciences, Kyoto Sangyo University, Kamigamo-motoyama, Kita-ku, Kyoto 603-8555, Japan
}

\date{Accepted XXX. Received YYY; in original form ZZZ}

\pubyear{2025}

\begin{document}
\label{firstpage}
\pagerange{\pageref{firstpage}--\pageref{lastpage}}
\maketitle

\begin{abstract}
The torus in Active Galactic Nuclei (AGN) is a complex dynamical structure of gas and dust. It is thought to be composed of an equatorial dusty disk and a polar dusty wind launched by radiation pressure. However, this picture is based on studies of moderately accreting AGN. Models suggest that the disk/wind structure will change with specific accretion rate. Here we examine the wind launching region in two high accretion rate objects, I Zw 1 (super-Eddington) and H0557-385 (high-Eddington), using high spatial resolution interferometric observations in the $K$-band from VLTI/GRAVITY and $LM$-bands VLTI/MATISSE. We recover wavelength-dependent sizes of the dust emission using a Gaussian and power law fit to the visibilities. Both objects are partially resolved and have radial sizes in the $KLM$-bands between 0.3 – 1.5 mas, with no signs of elongation. Combining our measurements with VLTI/MIDI $N$-band data gives a full multi-wavelength picture of the dust structure. We find that in H0557-385, the dust sizes between $3.5-8\:\mu\mathrm{m}$ are independent of the wavelength, roughly constant at $3-10$ sublimation radii. We argue that this indicates a direct view of the wind launching region and, together with an absence of polar elongation, this implies that any wind would be launched in a preferentially equatorial direction or blown out by strong radiation pressure. The size-wavelength relation for both objects shows a preferentially disky equatorial dust distribution. We conclude that there is strong evidence that the Eddington ratio shapes the inner dust structure, most notably the wind-launching region and wind direction.

\end{abstract}

\begin{keywords}
galaxies: active -- techniques: interferometric -- galaxies: nuclei -- galaxies: Seyfert
\end{keywords}



\section{Introduction}
Active galactic nuclei (AGNs) are some of the most powerful objects in the universe, yet much of their defining structure is highly concentrated to the central few parsecs of their host galaxies. These spatial scales have made it difficult to study the central engine (the supermassive black hole and the accretion disk) and its surrounding material. The dusty material is concentrated in the so-called dusty torus, the obscuring medium, a few tens of parsecs across. The gas and dust in this obscuring medium are a source of material for the accretion disk and provide a feedback connection to the rest of the galaxy. This obscuring medium was first proposed by \citet{antonucci1985} based on the idea that the viewing angle and an axisymmetric obscuring medium are responsible for the differences between Type 1 and Type 2 Seyferts. 
Observations show that the composition in the hotter, inner part of the torus is dominated by large graphite grains \citep[e.g.][]{kishimoto2007}. Assuming dust is accreted from the
host galaxy with an originally standard ISM dust composition, this implies that both grain size and composition will be radially stratified \citep{honig2019}.

High-resolution infrared interferometry has enabled us to directly resolve the dusty tori in several nearby galaxies. In particular, data from the former mid-infrared interferometer on the VLTI (Very Large Telescope Interferometer), MIDI (MID-infrared Interferometer, \citealt{leinert2003}), has indicated that this torus has a two-component structure: an equatorial disk and a polar outflow cone \citep[e.g.][]{honig2012, honig2013, burtscher2013,tristram2014, lopez-gonzaga2016}. Furthermore, infrared imaging analysis has shown the necessity for the existence of a dusty polar outflow \citep{honig2017,alonso-herrero2021,isbell2021}. 

Recently, the second generation VLTI instruments GRAVITY ($K$-band, \citealt{gravity2017}) and MATISSE ($L$, $M$, and $N$-band, \citealt{lopez2022}) have provided high quality data to enable the reconstruction of images of the central few parsecs in AGNs for the first time. The hottest dust near the sublimation radius (at temperatures of $\sim$1500 K), as traced by GRAVITY, is inferred to be showing an equatorially oriented ring in type 1 AGN \citep{gravity2020b}. MATISSE imaging of NGC 1068 and the Circinus Galaxy illustrates the multi-phase structure of the dust: the hotter dust imaged in the $L$- and $M$-bands, whilst still showing a polar extension, is significantly more concentrated and more luminous in the equatorial directions \citep{gamezrosas2022,isbell2023,leftley2024}. Finally, $N$-band data clearly shows the elongated polar structures in these objects. This implies that the hot, near-IR emitting dust is located close to the sublimation region in the plane of the accretion disk. In contrast, the cooler mid-IR dust is divided into two components, an equatorial disk and a polar outflow, with most of the flux located in the polar region \citep{honig2012,honig2019}.

High spatial resolution ALMA imagining has revealed dusty molecular tori that are preferentially aligned perpendicularly to the AGN axes \citep{combes2019,alonso-herrero2021,garcia-burrilo2021}. As these are disconnected from the galactic disk, they appear to be the larger scale (median radius $\sim 42\:\mathrm{pc}$) and lower temperature components of the equatorial dusty disk. In these structures, thermal dust emission is responsible for the majority of the $870\:\umu\mathrm{m}$ continuum flux. The same structure is found in a variety of molecular lines as well.

Radiation-hydrodynamic simulations have confirmed this arrangement. Radiation pressure launches winds at the inner edge of the dusty disk, forming a hollow polar cone \citep{wada2016,chan2016,chan2017,williamson2019}. Using detailed radiative transfer modelling of a multi-wavelength and resolution data set of the nucleus of the Circinus Galaxy, \citet{stalevski2019} managed to reproduce existing images with a dominant hollow polar cone. However, this is not the case for all AGNs – simulations and observations imply a combination of dust densities and AGN accretion rates, above which dust will be blown out before it can be launched in a polar wind \citep{ricci2017,venanzi2020,alonso-herrero2021,garcia-burrilo2021}. Increasing the accretion power at more moderate levels ($L/L_{Edd} < 0.1 $) will widen the opening angle of the cone, decreasing the prominence of the polar elongation \citep{williamson2020}. 

So far, the focus in interferometry studies of the polar winds has been AGNs in the mid-Seyfert regime ($L/L_\mathrm{Edd} \lesssim 0.05$). AGNs in the blowout region (see Fig. 4 in \citealt{alonso-herrero2021}), especially with high accretion rates, have been undersampled, due to the severe flux limitations in VLTI observations. These limitations have led us to speculate about dust structure in powerful AGNs. \cite{leftley2019} tentatively found that Eddington ratio increases, the resolved source fraction increases with respect to the unresolved source fraction in the $N$-band, where the polar outflows are strongest. This implies that in strong AGNs, more dust is blown into the dusty winds by radiation pressure. For the hot dust in the near-IR, \cite{gravity2020b} found that two luminous quasars with $L/L_\mathrm{Edd} \sim 1$ have sharply peaked dusty emission profiles and comparably small hot-dust sizes. This contrasts with the other, lower luminosity AGNs in their sample, which have more extended emission profiles and larger dust sizes. These are signs that the structure of the dust, including relative sizes and other parameters, evolves with Eddington ratio. With the new generation of VLTI instruments it is now possible to go fainter and expand our high-Eddington sample.

Studying the high end of the AGN accretion parameter space is important to constrain the physical mechanism that govern the dust structure in AGNs. It has been shown that the accretion rate has a large impact on the inner parts of AGN structure, in particular the nature of the accretion flow and observed SED shapes and outflow properties \citep[e.g.][]{czerny2003,temple2023}. This is arguably related with a change in underlying accretion physics at Eddington ratio of approximately $0.1-0.3$. Here, we are interested in tracing changes of the circumnuclear dusty structure with accretion rate and extend IR interferometric studies from the low Eddington regime at $L/L_\mathrm{Edd} < 0.3$ to high Eddington ratios $>0.3$. For this purpose we have observed two sources with high Eddington ratios.

I Zw 1 is the prototypical narrow line Seyfert 1 (NLS1), at a redshift of $\sim 0.061$ (255 Mpc) \citep{asmus2016}. Its black hole mass is estimated between $\log M_\mathrm{BH} \sim 6.97 - 7.16 $, based on reverberation mapping and line width results respectively \citep{hao2005,huang2019}. It is accreting at super-Eddington rates of $L/L_\mathrm{Edd} \sim 2.14$ \citep{hao2005}. This puts it very clearly into the blowout region shown in Fig. 4 in \citet{alonso-herrero2021}, making it an ideal object to study the importance of strong accretion. 

H0557-385 is also a Seyfert 1 galaxy, until now mostly studied in the X-ray, at redshift $z = 0.034$ and at a distance of 157 Mpc \citep{coffey2014,leftley2019}. The black hole has an intermediate mass, measured at about $\log M_\mathrm{BH} \sim 7.81$ using H$\beta$ line widths. With an accretion rate of $L/L_\mathrm{Edd} \sim 0.4$ it also belongs to the higher accreting objects \citep{coffey2014}. In addition, its infrared spectrum shows a significant bump in the $L$- and $M$-bands, which was not explainable using the MIDI $N$-band data \citep{kishimoto2011b}. We will investigate the source of this SED feature in this paper.

In this paper we investigate the dust structure in these two high-accretion AGNs, specifically with regards to the presence and/or orientation of the polar wind and the inner disk structure, using high angular resolution mid-IR interferometry data from MATISSE. Both AGNs have been previously observed in the $N$-band using MIDI, but failed to show any prominent elongations \citep{kishimoto2011b,burtscher2013,lopez-gonzaga2016}. GRAVITY data also exists for I Zw 1, which will provide a more coherent view of the multiphase structure of the dust. To examine the physical mechanisms in those sources, we also performed SED fits and produced model images using CAT3D-WIND to compare to our observations \citep{honig2017}. In Section \ref{data}, we will present our data acquisition and reduction, specifically with regards to MATISSE. In Section \ref{model}, we will explain our SED modelling procedure and our model image creation. In Section \ref{results}, we will collate our results, and compare our data with our mock observations of the model images. In Section \ref{discussion}, we will discuss these results in the context of prior interferometric studies of the dusty torus as well as what our modelling and observations tells us about the physical wind launching mechanism. 

\section{Data and Data Reduction}\label{data}

In the following section, we describe our data acquisition and reduction for the interferometric data. We also present the photometric and spectroscopic data we collected from the archives to build our SEDs. Finally, we discuss the geometric model that we used to derive sizes from the interferometric data.

\subsection{Interferometry}\label{data_int}

\subsubsection{MATISSE}\label{data_matisse}
MATISSE observations for I Zw 1 and H0557-385 were taken in the same night (24.09.2021), with the same settings, under program 0105.B-0346(A). Low resolution mode with UT baseline configuration was used to observe the faint dust continuum in the $L$- and $M$-bands, with a central wavelength of $3.5\:\umu\mathrm{m}$. $N$-band observations were also attempted but no fringes were detected, as they are below the current bias limit of MATISSE standalone observations \citep{lopez2022}.

Generally, A, G, and K stars are used to calibrate infrared interferometric data since their brightness in the optical gives a good AO performance. However, these stars are considerably bluer than AGNs, and the stars’ IR SED slopes are steeply declining. This will lead to flux at shorter wavelength to be overweighted, and in calibration this will shift the AGN slopes bluer. To correct for this, we instead used red M giants which have a more similar IR SED slope compared to AGNs. Stars were selected according to their V-K colours and small proper motions to weed out nearby red dwarf stars. 

\begin{figure*}
    \centering
    \includegraphics[width=0.49\textwidth]{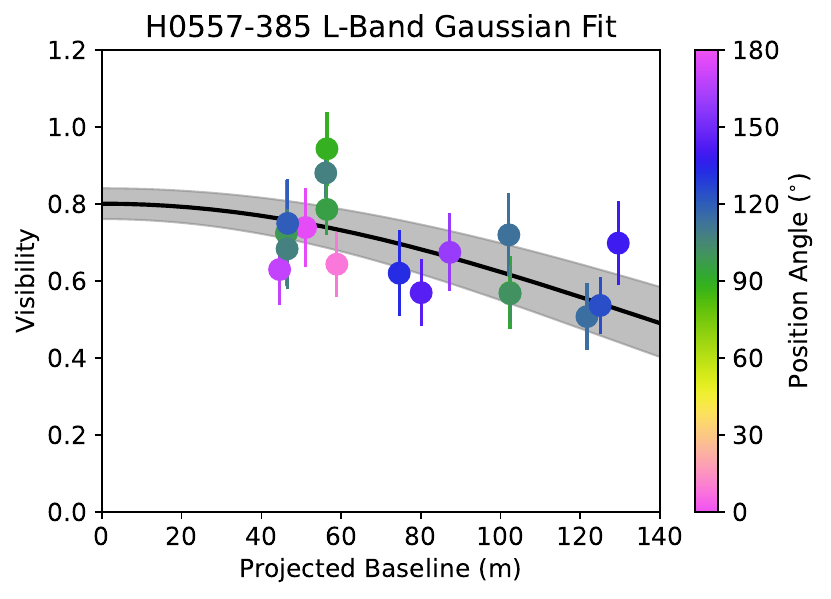}%
    \includegraphics[width=0.49\textwidth]{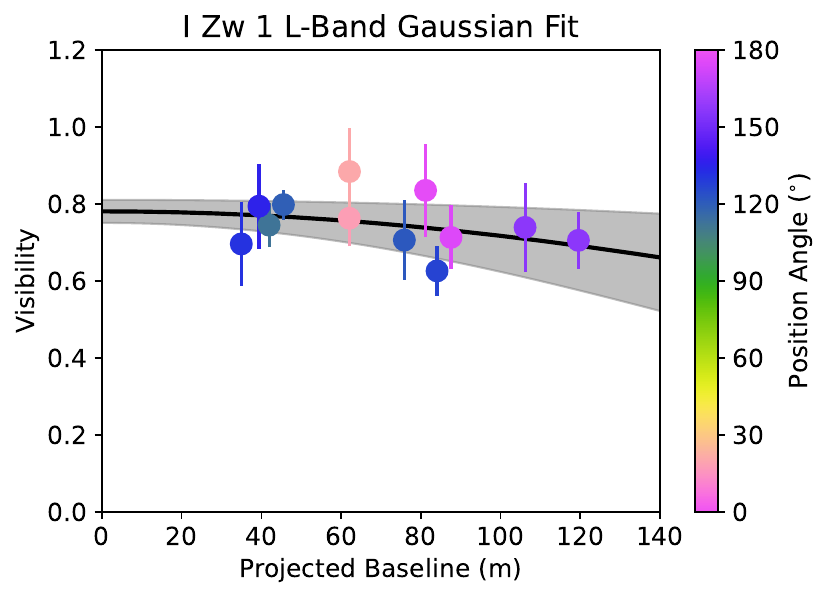}
    \caption{$L$-band visibilities and 1D Gaussian fit at $3.4\:\umu\mathrm{m}$ for H0557-385 (left) and I Zw 1 (right). Data is colored based on its position angle, and the 1D Gaussian fit according to Eq. \ref{eq:vis} is plotted with the black line, with the shaded region the error of the fit.}
    \label{fig:l_vis}
\end{figure*}

\begin{figure*}
    \centering
    \includegraphics[width=0.49\textwidth]{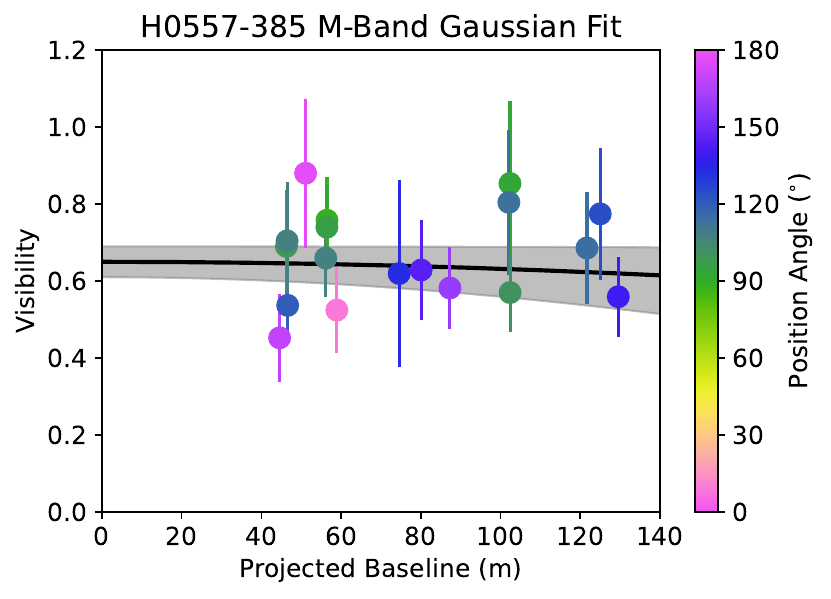}%
    \includegraphics[width=0.49\textwidth]{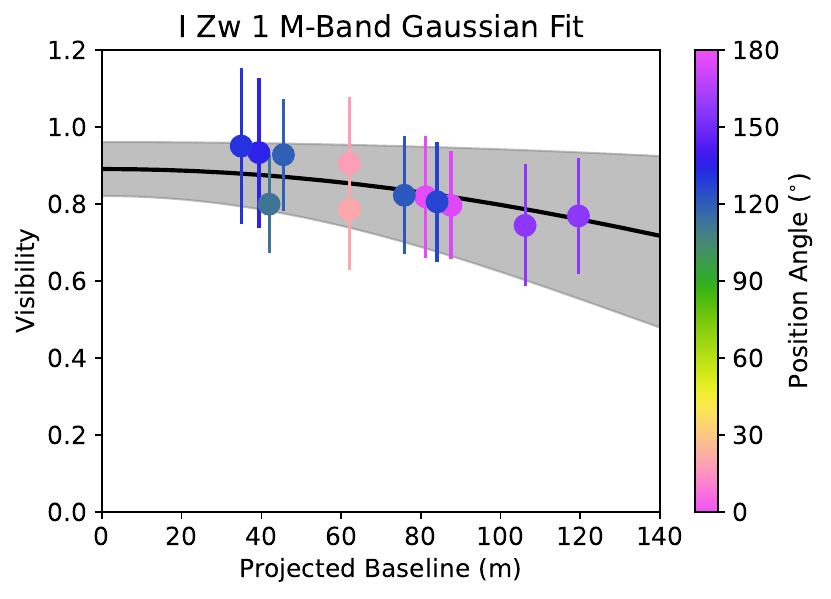}
    \caption{$M$-band visibilities and 1D Gaussian fit at $4.6\:\umu\mathrm{m}$ for H0557-385 (left) and I Zw 1 (right). Data is colored based on its position angle, and the 1D Gaussian fit according to Eq. \ref{eq:vis} is plotted with the black line, with the shaded region the error of the fit.}
    \label{fig:m_vis}
\end{figure*}

Data reduction was performed partly using version 1.7.6 of the MATISSE pipeline, and partly by using an algorithm developed by us specifically for faint objects. We used the MATISSE pipeline for a preliminary reduction, using its intermediate data products for our further analysis. First, we removed low signal-to-noise data, on a per wavelength bin per baseline per frame basis, essentially treating each baseline in each frame independently. This was done to keep as much data as possible. We then integrated over the entire band, since the SNR was too low to use the dispersed flux. The pipeline calculates the correlated flux in each baseline by taking the average over all frames\footnote{\href{https://ftp.eso.org/pub/dfs/pipelines/instruments/matisse/matisse-pipeline-manual-1.7.6.pdf}{MATISSE User Manual v1.7.6}}. However, since these are `faint’ AGNs -- close to or at the detection limit of MATISSE -- this will overestimate the correlated flux and visibility of the AGN. As can be seen in Figure \ref{fig:raw_dist}, the flux distributions (per frame) are clustered at low values but there is a long high flux tail. This tail will slew the average measurements. To mitigate this effect and determine the `true’ received interferometric flux from the science target, we deconvolved the detected flux distribution using the calibrator. As the calibrator’s frames are drawn from a different parent sample, we resampled it to match the science target frames. After frame by frame calibration (matching frames based on their position in the distribution), we successfully deconvolved the flux distribution, with a final correlated flux with errors of about 10\%. See Appendix~\ref{matisse data reduction} for further information. Finally, we calculated the visibility only using the chopped MATISSE photometry. The $L$-band data is shown in Figure \ref{fig:l_vis} and the $M$-band data is shown in Figure \ref{fig:m_vis}. 

Figures \ref{fig:l_vis} and \ref{fig:m_vis} show that both sources are partially resolved: the visibility decreases with increasing baseline length. Since they are not resolved completely, we are unable to do complex geometric modelling of the sources. Therefore, we focus on two main observables: elongations and sizes. The visibilities in Figures \ref{fig:l_vis} and \ref{fig:m_vis} are colour coded by position angle. As can be seen, no clear position angle dependency of the visibilities is present. To further test the presence of elongations, we modelled the source emission as a 2D elliptical Gaussian distribution. In this case, the real space flux distribution is given by
\begin{equation}
    F_\nu (x,y) = F_0 e^{-0.5 ((x\cos\theta+y\sin\theta)^2/\sigma_x^2 - (x\sin\theta-y\cos\theta)^2/\sigma_y^2)}
\end{equation}
where $F_0$ is a normalisation factor and $\theta$ is the position angle of the major axis of the ellipsis. In the case of elongation, we expect the angular sizes in the perpendicular directions, $\sigma_x$ and $\sigma_y$, to be significantly different. However, in modelling we found no constraints on $\sigma_x$, $\sigma_y$, and $\theta$. Therefore our data rules out any significant elongation in these objects. In addition, the closure phases are zero, implying an absence of any asymmetrical off-centre structure in the dust. Therefore, we interpret this as viewing a circular projected dust distribution and we can use one-dimensional (1D) Gaussian model with only two parameters to recover the brightness-weighted size of a partially resolved circular dust distribution. This is the same method to determine sizes as in \citet{gravity2020b,gravity2024}.


In a 1D Gaussian model, the visibility $V$ is given by
\begin{equation}
    V = V_0 e^{- 2\pi^2 B_{\mathrm{proj}}^2 \sigma^2 / \lambda^2} \label{eq:vis}
\end{equation}
\noindent where $V_0$ is the visibility normalisation, $B_{\mathrm{proj}}$ is the projected baseline length, $\lambda$ is the wavelength of the data, and $\sigma$ is the size of the Gaussian in radians, which we convert to mas \citep{gravity2024}. It should be noted that in this model, we do not force $V_0=1$. The single telescope fluxes may include contamination from astrophysical sources extending outside of the field-of-view of the interferometer, e.g. from nuclear stellar clusters, or the interferometer may suffer from instrumental losses, all leading to an effective $V_0<1$.

We used an MCMC method to fit the data with this model, with 16 walkers and 100,000 iterations each. As the burn-in, we discarded the first 5000 steps and thinned by 50\% to avoid autocorrelation. The complete set of results with one sigma errors is presented in Table \ref{tab:sizes_gauss}. For I Zw 1, we determine Gaussian sizes of $\sigma_L = 0.46 \pm 0.22$ mas and $\sigma_M = 0.71 \pm 0.41$ mas, for the $L$ and $M$-band emission, respectively. The $L$-band emission of H0557-385 is well constrained with a size of $\sigma_L = 0.79 \pm 0.11$ mas. In contrast, the $M$-band in H0557-385, is unconstrained, with a maximum size of $\sigma_{M,max} = 0.36 \pm 0.27$ mas. The fits are also shown in Figures \ref{fig:l_vis} and \ref{fig:m_vis}.

An alternative model to recover sizes from interferometric observation is adoption of a power law brightness distribution. Such a model is motivated by evaluating radiative transfer models and may have a more direct relation to physical brightness distributions \citep[e.g.][]{kishimoto2007,honig2010,kishimoto2011b}. However, such models require assuming zero baseline visibility $V_0 = 1$ as the anchor due to the lack of sufficiently high quality data to recover the baseline-dependent visibility slope without this assumption. Astrophysical contamination of the single-telescope flux will be significantly less at longer wavelengths \citep[and effectively absent in the $N$-band, see][]{kishimoto2011b,asmus2016}. As such, we consider the power law modelling approach a reasonable alternative to the Gaussian model in the $L$- and $M$-bands.

\begin{figure*}
    \centering
    \includegraphics[width=0.49\textwidth]{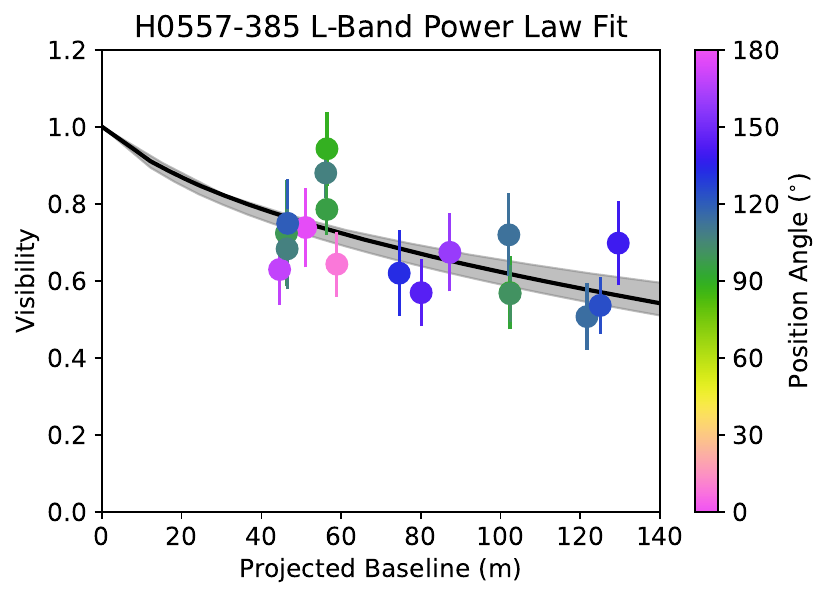}%
    \includegraphics[width=0.49\textwidth]{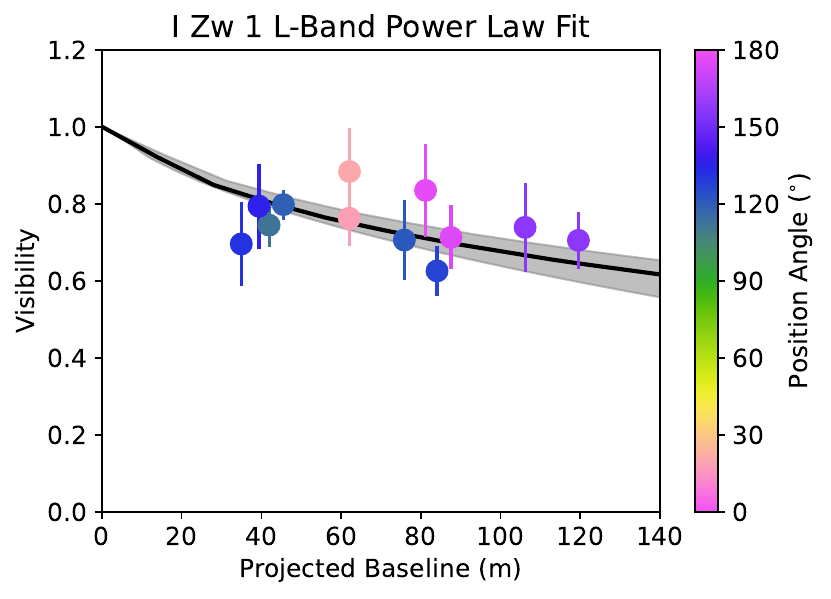}
    \caption{Power law fit of the $L$-band visibilities at $3.4\:\umu\mathrm{m}$ for H0557-385 (left) and I Zw 1 (right). Data is colored based on its position angle, and the power law fit is plotted with the black line, with the shaded region the error of the fit.}
    \label{fig:l_vis_pl}
\end{figure*}

\begin{figure*}
    \centering
    \includegraphics[width=0.49\textwidth]{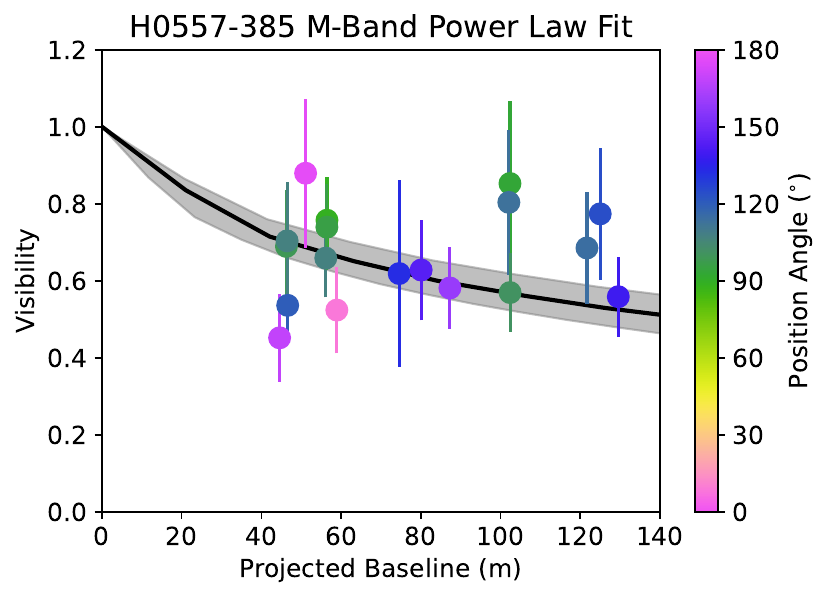}%
    \includegraphics[width=0.49\textwidth]{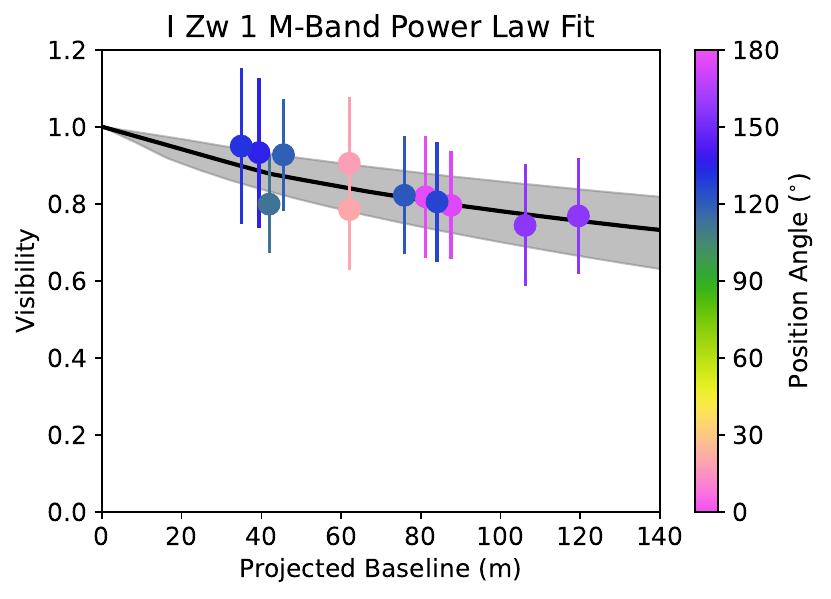}
    \caption{Power law fit of the $M$-band visibilities at $4.6\:\umu\mathrm{m}$ for H0557-385 (left) and I Zw 1 (right). Data is colored based on its position angle, and the power law fit is plotted with the black line, with the shaded region the error of the fit.}
    \label{fig:m_vis_pl}
\end{figure*}

For the power law model, we created a grid of $10000 \times 10000$ pixels and fill it with a circosymmetric brightness distribution, with a central hole. The brightness distribution follows $r^{-\alpha}$ where $\alpha$ is the power law index and $r$ the distance from the centre. The radius of the central hole is set by the sublimation radius $r_\mathrm{in}$. Visibilities are extracted from the Fourier transformed grid, normalised such that $V_0 = 1$. In this model, $\alpha$ and $r_{in}$ are the fitting parameters. We fitted the data by minimising the $\chi^2$. It should be noted that $r_\mathrm{in}$ is not well constrained in this model as the observed baseline lengths do not resolve the corresponding spatial scales. Therefore, we calculate the best fit power law index marginalised over $r_\mathrm{in}$, shown in Table \ref{tab:sizes_pl} and provide with one sigma (68\%) confidence intervals. For I Zw 1 the marginalised power law index in the $L$-band is $\alpha = 2.48 \pm 0.12$. In the $M$-band, $\alpha = 2.60 \pm0.38$. For H0557-385 the $L$-band marginalised power law index was found to be $\alpha = 2.40 \pm 0.10$, and in the $M$-band $\alpha=2.22 \pm 0.13$. In addition, we evaluate the one sigma $\chi^2$ contours over all model parameters to find the uncertainty range for our models. These are plotted together with best fit models in Figures \ref{fig:l_vis_pl} and \ref{fig:m_vis_pl}.


\subsubsection{GRAVITY}\label{data_grav}

\begin{figure}
    \centering
    \includegraphics[width = 0.49\textwidth]{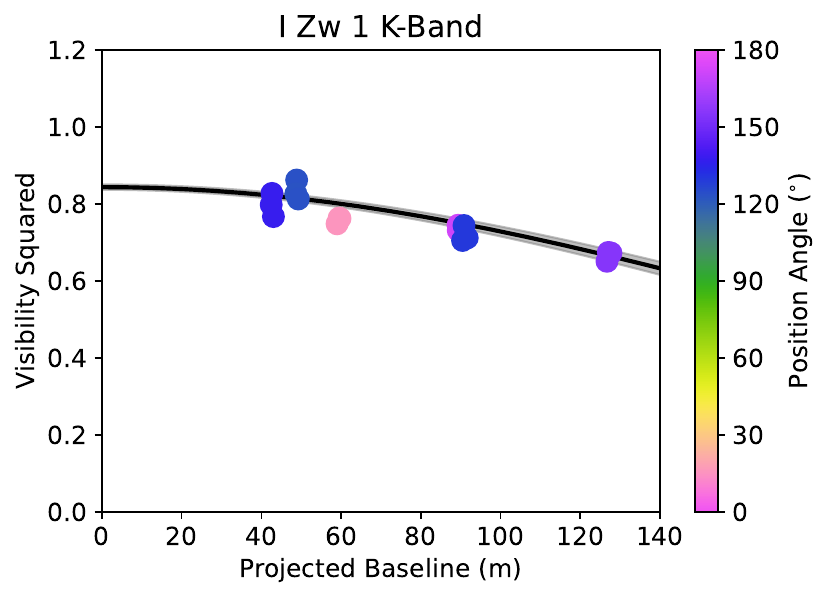}
    \caption{$K$-band squared visibilities at $2.2 \: \umu\mathrm{m}$ for I Zw 1. Data is colored based on its position angle, and the 1D Gaussian fit according to Eq. \ref{eq:vis} is plotted with the black line, with the shaded region the error of the fit. Errors are too small to be visible.}
    \label{fig:grav_vis}
\end{figure}

GRAVITY data is only available for I Zw 1, taken as part of program 1103.D-0626(C) on 25.07.2021. It was observed in dual-field off-axis mode with low spectral resolution and UT baseline configuration, using the fringe tracker to record the fringes using 100\% of the light, while the science combiner was pointing at the sky. For data reduction, we used version 1.4.2 of the GRAVITY pipeline. In addition, according to \cite{gravity2020b}, we selected frames with group delays of $< 3\: \umu\mathrm{m}$ to account for visibility losses due to the atmosphere. The visibilities squared are plotted in Figure \ref{fig:grav_vis}. In this case, we constrained our models to the 1D Gaussian model. The dust that dominates $K$-band emission is emitted within a small radius. In this case, the size measured by the Gaussian half light radius is a good approximation of the power law half light radius \citep{kishimoto2011b}. Like for the $L$ and $M$ bands, we cannot make the assumption that $V_0 = 1$, due instrumental effects and contaminations, which are significant in the $K$-band \citep{gravity2020b}. We used the same 1D Gaussian to fit the data as we did for MATISSE data (Eq. \ref{eq:vis}), to estimate the maximum size of the hot dusty region, using $V^2$ instead of $V$. However, the tail end of the accretion disk emission contributes significantly to the near-IR luminosity. To estimate this, we fit $F_\nu \propto \nu^{1/3}$ to the optical flux of I Zw 1 as shown in the SED in Figure \ref{fig:sed}. In the $K$-band, the point source contribution is $f_{pt} = 0.07$. This agrees well with the range of values of $5-25\%$ typically found in type 1 AGN as measured by \cite{kishimoto2007}. With this, we use Eq. 5 from \cite{gravity2020b} to account for the point source contribution. We get a final size of the partially resolved hot dust region in the $K$-band in I Zw 1 of $\sigma_K = 0.29\pm0.01$ mas, with $V_0^2 = 0.843 \pm 0.007$. Results are summarised in Table \ref{tab:sizes_gauss}.

\begin{table*}
    \centering
    \begin{tabularx}{\textwidth}{YYYYYY}
         Object & Wavelength & $V_0$ &Angular Size $\sigma$ (mas) & $R_{1/2}$ (pc) & $R_{1/2}$ ($R_{\mathrm{sub}}$)$^c$\\
         \hline
         H0557-385 & $3.4\:\umu\mathrm{m}$ & $0.80\pm0.04$ &$0.79\pm0.11$ & $0.71\pm0.10$ & $5.9\pm0.8$ \\
         & $4.6\:\umu\mathrm{m}$ & $0.65\pm0.04$&$\le 0.36\pm0.27$ & $\le 0.32\pm0.24$ & $\le 2.7\pm2.0$ \\
         & $8.5\:\umu\mathrm{m}$ & & & $0.77_{-0.29}^{+0.48,a}$ & $6.4^{+4.1}_{2.4}$ \\
         & $12.4\:\umu\mathrm{m}$ & & $18^{+15,b}_{-4.7}$ & $<2.4^{b}$ & $<20.0$ \\
         & $13.0\:\umu\mathrm{m}$ & & & $1.89\pm_{-0.72}^{+1.17,a}$ & $15.8_{-6.0}^{9.8}$ \\
         I Zw 1 & $2.2\:\umu\mathrm{m}$ & $0.918\pm0.004$ &$0.29\pm0.01$ & $0.42\pm0.01$ & $2.3\pm0.06$ \\
         & $3.4\:\umu\mathrm{m}$ & $0.78\pm0.03$ & $0.46\pm0.22$ & $0.67\pm0.32$ & $3.7\pm1.6$ \\
         & $4.6\:\umu\mathrm{m}$ & $0.89\pm0.07$ & $0.71\pm0.41$ & $1.03\pm0.60$ & $5.7\pm3.3$ \\
         & $12.7\:\umu\mathrm{m}$ & & $10.2^{+8.5,b}_{-6.4}$ & $< 4.01^{b}$ & $<22.5$ \\
         \vspace{-0.3cm}\\
         \hline
         \multicolumn{6}{l}{\textbf{Notes.}} \\
         \multicolumn{6}{l}{$^a$ Sizes from \citet{kishimoto2011b}, measured using a power law and corrected with a factor of 1.5.} \\
         \multicolumn{6}{l}{$^b$ Sizes from \citet{burtscher2013}, measured using a point source and a Gaussian, which introduces upper limits for $R_{1/2}$} at point source fractions above 0.5.\\ 
         \multicolumn{6}{l}{Corrected with a factor of 1.5.}\\
         \multicolumn{6}{l}{$^c$ Sublimation radii used are $R_{\mathrm{sub}} = 0.12\:\mathrm{pc}$ for H0557-385 and $R_{\mathrm{sub}} = 0.18\:\mathrm{pc}$ for I Zw 1 (see Section \ref{lbol}).}
    \end{tabularx}
    \caption{The Gaussian fit results to the interferometry ($V_0$ and $\sigma$) and additional sizes of the objects at different wavelengths. The angular size is the $\sigma$ of the Gaussian fit to the visibility data, in the $K$-band corrected for the accretion disk contribution. The physical radii are the half light radii (here, the HWHM).}
    \label{tab:sizes_gauss}
\end{table*}

\begin{table*}
    \centering
    \begin{tabularx}{\textwidth}{YYYYYY}
         Object & Wavelength & Power Law Index $\alpha$ & $R_{1/2}$ (mas) & $R_{1/2}$ (pc) & $R_{1/2}$ ($R_{\mathrm{sub}}$)$^c$\\
         \hline
         H0557-385 & $3.4\:\mu\mathrm{m}$ &$2.40\pm0.10$&$0.87\pm0.19$ & $0.66\pm0.14$& $5.5\pm1.2$ \\
         & $4.6\:\mu\mathrm{m}$ & $2.22\pm0.13$&$1.45\pm0.38$ & $1.10\pm0.29$ & $9.2\pm2.4$ \\
         & $8.5\:\mu\mathrm{m}$ & & & $0.51_{-0.19}^{+0.32,a}$ & $4.3_{-1.6}^{+2.7}$ \\
         & $12.4\:\umu\mathrm{m}$ & & $18^{+15,b}_{-4.7}$ & $<1.6^{b}$ & $<13.3$ \\
         & $13.0\:\mu\mathrm{m}$ & & & $1.26_{-0.48}^{+0.78,a}$ & $10.5_{-4.0}^{+6.5}$ \\
         I Zw 1 & $3.4\:\mu\mathrm{m}$ & $2.48\pm0.12$ & $0.73\pm0.19$ & $0.90\pm0.23$ & $5.0\pm1.3$ \\
         & $4.6\:\mu\mathrm{m}$ & $2.60\pm0.38$ & $0.58\pm0.29$ & $0.72\pm0.36$ & $4.0\pm2.0$ \\
         & $12.7\:\mu\mathrm{m}$ & & $10.2^{+8.5,b}_{-6.4}$ & $< 2.7^{b}$ & $<15.0$ \\
         \vspace{-0.3cm}\\
         \hline
         \multicolumn{6}{l}{\textbf{Notes.}} \\
         \multicolumn{6}{l}{$^a$ Sizes from \citet{kishimoto2011b}, measured using a power law.} \\
         \multicolumn{6}{l}{$^b$ Sizes from \cite{burtscher2013}, measured using a point source and a Gaussian, which introduces upper limits at point source fractions above 0.5.}\\
         \multicolumn{6}{l}{$^c$ Sublimation radii used are $R_{\mathrm{sub}} = 0.12\:\mathrm{pc}$ for H0557-385 and $R_{\mathrm{sub}} = 0.18\:\mathrm{pc}$ for I Zw 1 (see Section \ref{lbol}).}
    \end{tabularx}
    \caption{The power law fit results to the interferometry and additional sizes of the objects at different wavelengths. The power law is characterised by its power law index $\alpha$, where the brightness is radially distributed according to $r^{-\alpha}$, and the sublimation radius $r_\mathrm{in}$. The power law index here shown is marginalised over $r_\mathrm{in}$.} All radii are half light radii, measured at $V=0.5$. $K$-band results are not included as due to the small physical size of the region, it is approximated well using a Gaussian.
    \label{tab:sizes_pl}
\end{table*}

\subsection{SEDs}

For this work, we collated IR SED data from archival sources to construct the SED of the dusty `torus' and to then model the SED using CAT3D-WIND. We focus on the IR emission, in the range from $1\:\umu\mathrm{m}$ to $\sim 100\:\umu\mathrm{m}$. This covers emission originating predominantly from within the central $\sim$ 100 pc \citep{asmus2014}, which corresponds to an angular size of $\sim 0\farcs{0}{8} - 0\farcs{1}{3}$ for our objects. These resolutions are very difficult to attain with single dish photometric and spectroscopic observations. We aim for higher resolutions to examine galactic contributions such as starbursts, especially in larger aperture measurements. From this, we can judge to what extent the SED reflects the dusty `torus' emission and the applicability of SED modelling.

\begin{figure*}
    \centering
    \includegraphics[width=0.49\textwidth]{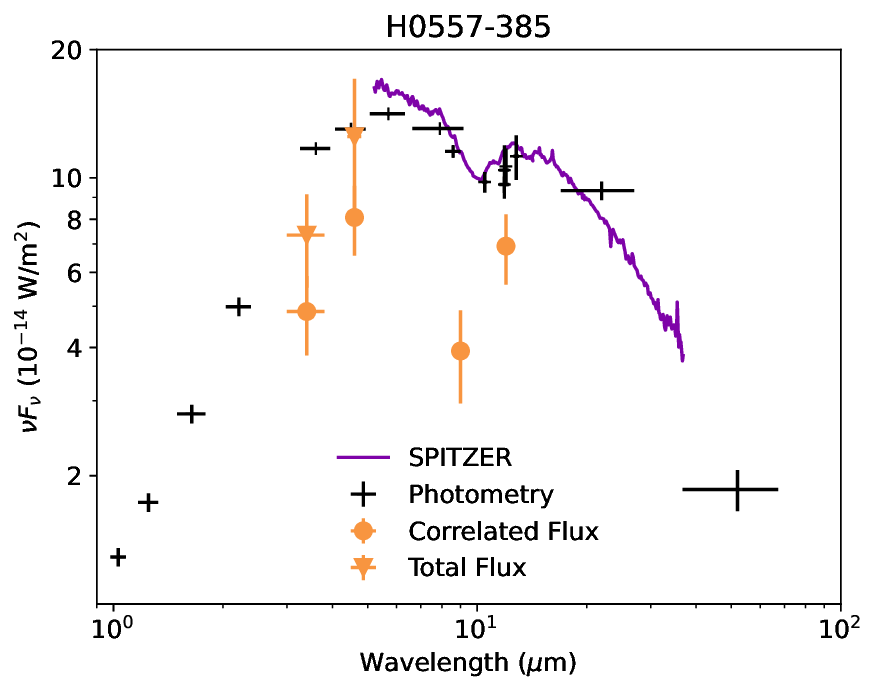}%
    \includegraphics[width=0.49\textwidth]{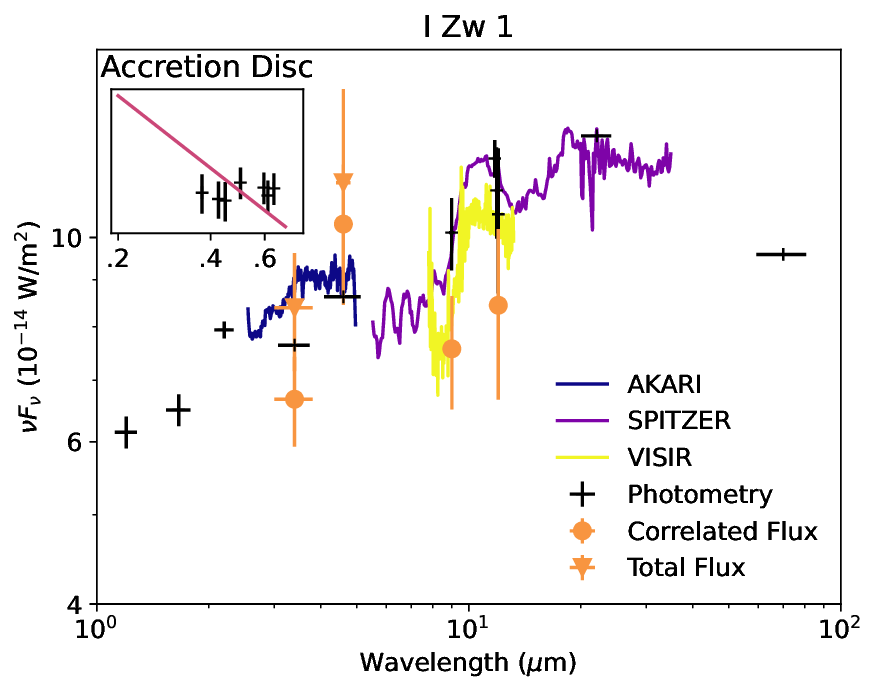}
    \caption{SEDs for H0557-385 (left) and I Zw 1 (right), with interferometric measurements from MATISSE in the $L$- and $M$-bands, and from MIDI in the $N$-band overplotted in orange circles. Higher circles are total flux measurements and lower circles are correlated flux measurements, except in the $N$-band where only correlated flux measurements exist.}
    \label{fig:sed}
\end{figure*}

For H0557-385, we have compiled a set of high resolution photometry from $Y$-band to $N$-band, taken with the explicit goal of studying the dust in the AGN \citep{kishimoto2011b,asmus2014}. Furthermore, we have included two long wavelengths measurements from IRAS to characterise the behaviour of the colder dust \citep{moshir1990}. Due to the large aperture, these can be seen as strict upper limits. Finally, we have overlaid the SPITZER point source spectrum in Figure \ref{fig:sed} to clearly illustrate the unusual mid-IR features. These include an unusually high $LM$-band bump and a $10\:\umu\mathrm{m}$ silicate feature in absorption – in a Type 1. However, the Spitzer data will not be further used in the analysis since the aperture is $\gtrsim 3\times$ the AGN infrared emitting region. Nevertheless, we do not expect significant starburst activity since the star formation rate is $\lesssim 1.6\:\mathrm{M_{\odot}yr^{-1}}$ \citep{shimizu2017}. Regarding variability, while the X-ray emission has seen significant flux variability due to absorption events, during these times optical emission has stayed nominal \citep{coffey2014}. Full SED data can be found in Table \ref{tab:data_h0557-385}, and the SED is displayed in Figure \ref{fig:sed}. 

I Zw 1 is a more complex case due to the presence of a nuclear starburst and near-IR accretion disk contribution \citep{schinnerer1998,kishimoto2007}. Using ALMA data, \cite{fei2023} showed that the starburst is within the central 1 kpc of the galaxy ($0\farcs{8}$). On the other hand, the VISIR spectrum of I Zw 1 does not show significant starburst emission, which implies that only apertures $\gtrsim 0\farcs{4}$ contain significant contamination from star formation \citep{jensen2017}. Consequently, as illustrated in Table \ref{tab:data_izw1} with the extraction aperture sizes, the majority of the available data is contaminated by starburst emission. We have amassed data from $\sim 1\:\umu\mathrm{m}$ to $70\:\umu\mathrm{m}$, both spectroscopic and photometric, that include starburst SED signatures \citep{hickox2018}. Starbursts predominantly emit in the mid-IR, overlapping directly with the dusty torus emission. In addition, starformation has an irregular shape with prominent emission features. The emission lines can be neglected because we are dealing with broadband SED fitting and not spectral fitting. Adding a starburst SED's sub-$10\:\umu\mathrm{m}$ emission, which peaks in the $L$-band region, will offset the balance between the sub-$10\:\umu\mathrm{m}$ and $\gtrsim10\:\umu\mathrm{m}$ emission by overestimating the sub-$10\:\umu\mathrm{m}$ contribution \citep{hickox2018}. This makes it very difficult to model the underlying torus emission accurately, except through high spatial resolution measurements particularly in the $LM$-band region.

The near-IR emission will be impacted by the tail of the accretion disk. Crucially, this will impact the $K$-band size of the dust as measured with GRAVITY. To estimate the accretion disk contribution used in Section \ref{data_grav}, we obtained continuum fluxes from a UVES spectrum (ID 085.C-0172). The continuum is relatively flat, which matches with previous optical spectra \citep[e.g. ][]{baldwin2004}. We do have high resolution $N$-band coverage, both in photometry and spectroscopy, which is dominated by dusty ‘torus’ emission \citep{asmus2014,jensen2017}. Comparing the high resolution VISIR spectrum to the low resolution SPITZER spectrum, in the mid-IR, we see not only a difference in flux levels but also a change in the spectral shape especially around the silicate feature at $\sim 10\:\umu\mathrm{m}$. Variability is not a large concern as the ASAS-SN light curve from $2012 - 2019$ shows long term flux stability \citep{huang2019}. In the short term (cadence of several days), variability is unimportant, with the optical flux varying between 3--9\% \citep{huang2019}. We note that our interferometric flux in the $M$-band is larger than the AKARI spectrum, which is likely due to flux calibration issues. Full SED data can be found in Table \ref{tab:data_izw1}, and the SED is displayed in Figure \ref{fig:sed}.

\section{CAT3D-WIND}\label{model}

In this section, we simultaneously model the SEDs and visibilities. We focus particularly on whether there is a polar wind present. SED modelling alone is insufficient to reveal structural information due to degeneracies \citep{feltre2012}. Therefore, we test our best fit SED models against visibilities to recover information about structure.

We use the radiative transfer code CAT3D-WIND to model the dust structure \citep{honig2010,honig2017}. This model assumes a two-phase disk+wind structure, which has been inspired by IR interferometry of a set of AGN and matches the geometric distribution obtained results from radiation-hydrodynamic simulations. In this model, the wind is described as a hollow polar cone with a mass distribution independent of the disk \citep{wada2016,williamson2020}. The dust is contained in randomly distributed clouds, assuming a clumpy torus. Through Monte Carlo and radiative transfer and ray tracing, the SEDs for different parameter values are extracted. In addition, images of the torus at different wavelengths can be obtained. The model parameter that is of specific interest to us here is the wind to disk ratio, i.e. the amount of dust contained within the wind versus within the disk. Furthermore, the wind opening angle is of interest. Using this model enables us to compare our observations to a realistic representation of the dust structure, which includes the wind (which can also be switched off). Specifically, we are able to examine the importance of the wind in comparison to the disk and how large the wind opening angle is. The provision of images allows us to extract interferometric information from the model to compare to our observations. 

\begin{figure}
    \centering
    \includegraphics[width=0.49\textwidth]{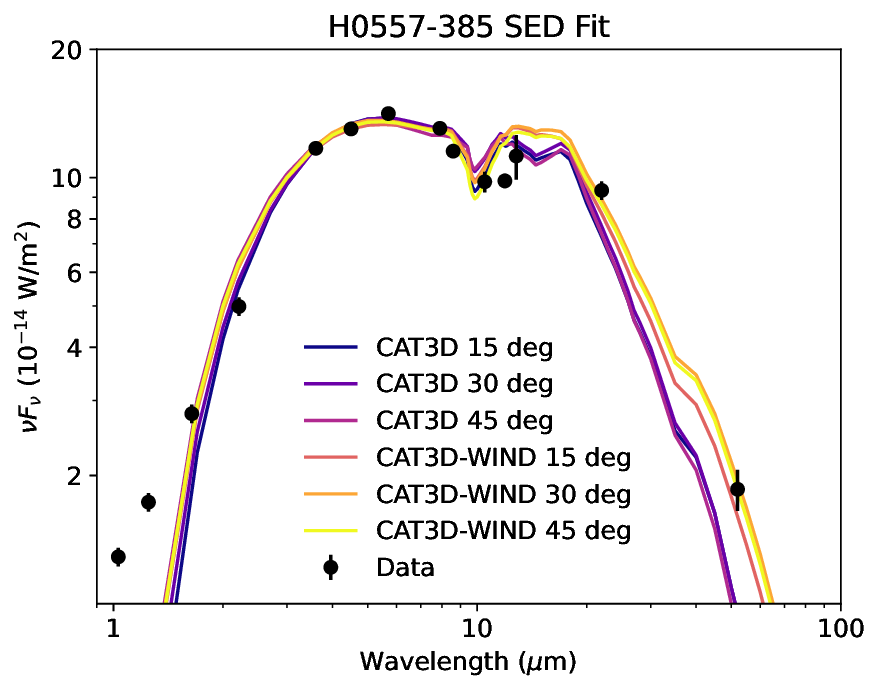}
    \caption{Best-fit SED fits for H0557-385, without and without wind, over inclinations of 15, 30, and 45 degrees. The data is plotted with the black dots. Table \ref{tab:mod_res} contains the best-fit model parameters.}
    \label{fig:sed_fit}
\end{figure}

\subsection{H0557-385}

\begin{figure*}
    \centering
    \includegraphics[width=0.49\textwidth]{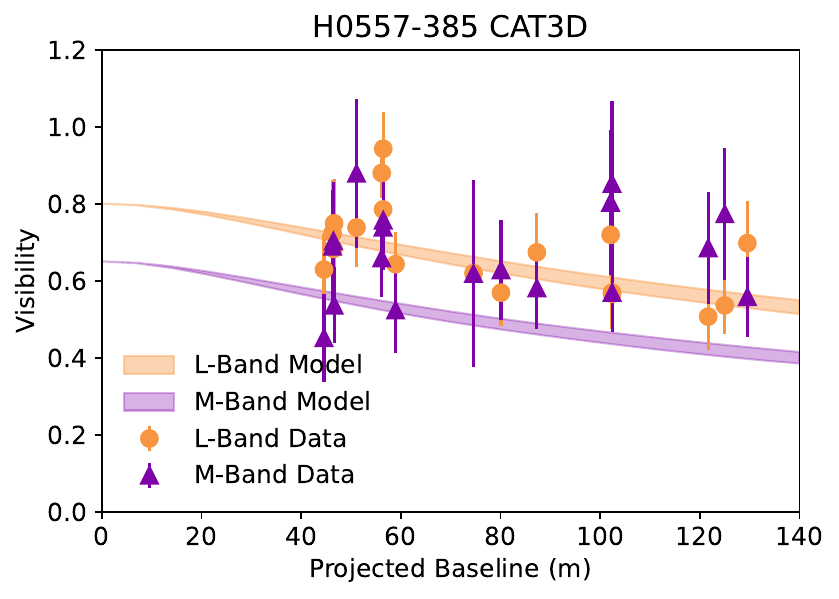}%
    \includegraphics[width=0.49\textwidth]{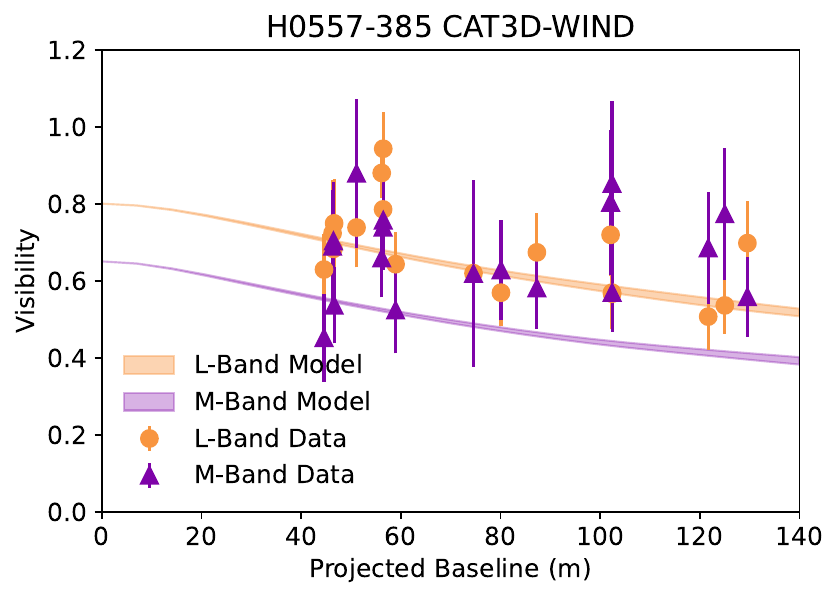}
    \caption{Interferometric results for the best fit models for H0557-385 with our observations overlaid. The shaded regions of the models cover the visibilities for all position angles. The $L$-band is in orange and the $M$-band is in purple. On the left is the best fit model without wind and on the right is the best fit model with wind. }
    \label{fig:model_obs}
\end{figure*}

Examining the SED of H0557-385 (Figure \ref{fig:sed}), we can clearly see a prominent silicate absorption feature around $10\:\umu\mathrm{m}$. However, in Type 1s, such as H0557-385, we expect to see the silicate feature in emission \citep{honig2010}. The host galaxy of this source is highly inclined. Therefore, the silicate absorption is probably of host-galactic origin, and we included ISM extinction in our SED fitting to account for the deep absorption feature \citep{chiar2006, goulding2012}. For modelling, we used the photometric data as seen in Figure \ref{fig:sed} and Table \ref{tab:data_h0557-385}. We combined the three photometric points around $12\:\umu\mathrm{m}$ into one, so as to not overweight the fit in that region. We added constraints on some model parameters as fitting these models unconstrained does not lead to conclusive results. As we know this is a Type 1, we restricted possible inclinations to below 60 degrees. From \cite{rokaki1999} we can also presume that the torus is probably not viewed face on, likely between $20 - 40$ degree. Therefore, we limit inclination angles to $15 - 45$ degree. We restricted the value of the optical depth $\tau$ such that the silicate feature is in absorption. All fit results are shown in Table \ref{tab:mod_res} and Figure \ref{fig:sed_fit}. A host galaxy extinction (Figure \ref{fig:sed_fit}) improves the shape of the fit, reproducing the SED shape. However, through introducing degeneracies, constraints on parameters become less tight. Goodness-of-fit values for the CAT3D and CAT3D-WIND fits do not differ significantly. In addition, as can be seen in Figure \ref{fig:sed_fit}, there is very little difference between the fits. Essentially, simply adjusting the amount of extinction leads to a viable fit. At the longest wavelengths, the data does prefer the wind-added model. The reason for this is that the modelled wind has a flatter dust distribution than the disk. Therefore, at larger distances, there is comparably more dust in the wind than at the same distance in the disk. This means that predominantly cooler, longer-wavelength emission at those distances will be dominated by the wind.

Furthermore, we created images of the torus based on the best fit models. We then mock observed these with a set of ideal $uv$-points, maximising the coverage in all directions. Our aim here is to examine how strongly the emission is elongated for these models. Further, we want to see whether this elongation could be detected with interferometry. The results, overlaid with the observations, are shown in Figure \ref{fig:model_obs}. It appears that polar structure is naturally suppressed. In addition, for inclinations of Type 1s, it is not sufficiently detectable in this object with current interferometric instrumentation -- a visibility accuracy of the order of 1\% would be needed. 

\subsection{I Zw 1}

The complexity of the SED of I Zw 1 makes a fit with a torus model very ambiguous and hence does not provide any meaningful constrains on the dust distribution. Specifically, the shape at $\lesssim 10\:\umu\mathrm{m}$ is badly defined, in contrast to the $\gtrsim 10\:\umu\mathrm{m}$ emission. Since this is not the focus of this work, we have decided not to pursue this problem further. 

\section{Results}\label{results}

\subsection{Bolometric Luminosity and the Sublimation Radius}\label{lbol}
Our goal is to compare the interferometric sizes between our two objects, as well as simple dust emission models and objects with lower Eddington ratios. Through this, we can look at the influence of the Eddington ratio on the structure of the torus. However, the physical radii are to the first degree dependent on the bolometric luminosity of the central engine. To account for this, we scale the interferometric sizes by the sublimation radius, removing any bolometric luminosity dependencies \citep{gravity2020b}. We hence first determine the bolometric luminosities of our objects. There are a variety of techniques to find $L_\mathrm{bol}$, e.g.\ based on the mid-IR luminosity, $\lambda L_{5100\textnormal{\AA}}$, or X-ray luminosities, especially $14-195$ keV. Here, our goal is to determine $L_\mathrm{bol}$ in the most self-consistent way for maximum comparability, rather than the intrinsic value. To achieve this, we collected $L_{\mathrm{MIR}}$ and $L_{2-10\:\mathrm{keV}}$ for both objects, and $L_{14-195\:\mathrm{keV}}$ for H0557-385, from \citet{asmus2015}. The intrinsic I Zw 1 $5100\textnormal{\AA}$ luminosity is taken from \citet{huang2019} and for H0557-385 from \citet{coffey2014}. 
From the hard X-ray and optical luminosities we found the bolometric luminosity directly through established relations. For the $L_{2-10\:\mathrm{keV}}$ and $L_{\mathrm{MIR}}$, we first transformed these into a hard X-ray luminosity, and then calculated the bolometric luminosity based on that. Our method is based on Appendix A of \citet{gravity2020b} and the luminosity relations used therein: the $L_\mathrm{bol} - L_{14-195\:\mathrm{keV}}$ relation from \citet{winter2012}, $L_\mathrm{bol} - \lambda L_{5100\textnormal{\AA}}$ from \citet{trakhtenbrot2017}, and $L_{14-195\:\mathrm{keV}} - L_{2-10\:\mathrm{keV}}$ from \citet{winter2009}. The $L_{2-10\:\mathrm{keV}} - L_{\mathrm{MIR}}$ relation is taken from \citet{asmus2015}. 

In I Zw 1, we find a spread in $L_\mathrm{bol}$ of 1 dex, giving a factor of 3 for the uncertainty in $R_{\mathrm{sub}}$. This is mainly driven by its relative faintness in the $2-10$ keV band in comparison to its mid-IR luminosity \citep{asmus2015}. In general, I Zw 1 exhibits highly complex behaviours in the X-ray, including reflections from behind the black hole and ultra-fast outflows \citep{wilkins2021,rogantini2022}. This implies that the X-ray based luminosity is not reliable in this object.

On the other hand, H0557-385 is far more compact in the spread in $L_\mathrm{bol}$, $\sim 0.5$ dex. Again, X-ray based luminosities are assumed not to be particularly reliable in this object due to a peculiar absorption event in the 2000s, that solely impacted the X-ray emission but not the optical \citep{coffey2014}.

As the $ \lambda L_{5100\textnormal{\AA}}$ derived bolometric luminosities approximately cover the center of both distributions, we have decided to use these here. For I Zw 1, the intrinsic $\lambda L_{5100\textnormal{\AA}} = 10^{44.50}$ erg/s, which gives $L_\mathrm{bol} = 10^{45.36}$ erg/s \citep{trakhtenbrot2017,huang2019}. For H0557-385 , $\lambda L_{5100\textnormal{\AA}} = 10^{44.12}$ erg/s, giving $L_\mathrm{bol} = 10^{45.01}$ erg/s \citep{coffey2014}. To estimate the sublimation radius we used the relation between $R_{\mathrm{sub}}$ and $L_\mathrm{bol}$ as presented in \cite{gravity2020b} (see their Figure 7 and Section 5.2 for details). In I Zw 1, $R_{\mathrm{sub}} = 0.18\:\mathrm{pc}$, and in H0557-385 $R_{\mathrm{sub}} = 0.12\:\mathrm{pc}$. This bolometric-luminosity-derived sublimation radius of H0557-385 is very consistent with sublimation radii derived from SED fitting, $0.1-0.12$ pc (Table \ref{tab:mod_res}). This corroborates our usage of $\lambda L_{5100\textnormal{\AA}}$ derived bolometric luminosities. 

\subsection{Interferometric Sizes}\label{int_size}
With a large wavelength coverage of interferometric sizes from GRAVITY, MATISSE, and MIDI, we can examine the radial distribution of the dust, assuming the $K$-, $LM$- and $N$-band emission mainly trace the hot ($T > $1000K), warm ($T \sim$ 600K) and cool ($T \sim$ 300K) dust respectively. This will give us important insights into the structure of the dusty disk in these high accretion objects. For comparable size estimates of these emission regions we determined their half light radii $R_{1/2}$, which includes half of the emission within it. For the Gaussian sizes $\sigma$ we obtained by fitting the interferometric data in Section \ref{data_int}, we determined the half width at half maximum (HWHM $= 1.1775\sigma$). This is the half light radius for a radially symmetric 2D Gaussian model. Resulting half light radii in both pc and in units of the sublimation radius $R_{\mathrm{sub}}$ can be seen in Table \ref{tab:sizes_gauss}. Since the data quality for H0557-385 in the $M$-band is so low, we have only managed to obtain an estimated upper limit. For our power law model fits, we determined the half light radius $R_\mathrm{1/2} = R_{V=0.5}/4.5$ with $R_{V=0.5}$ as the spatial wavelength at $V=0.5$ \citep[see][for details]{kishimoto2011b}. The uncertainties are derived from the one sigma contours seen in Figure \ref{fig:l_vis_pl} and \ref{fig:m_vis_pl} and discussed in Section \ref{data_matisse}. These are presented in Table \ref{tab:sizes_pl}, in units of mas, pc, and scaled by the sublimation radius $R_{\mathrm{sub}}$.

\begin{figure}
    \centering
    \includegraphics[width = 0.49\textwidth]{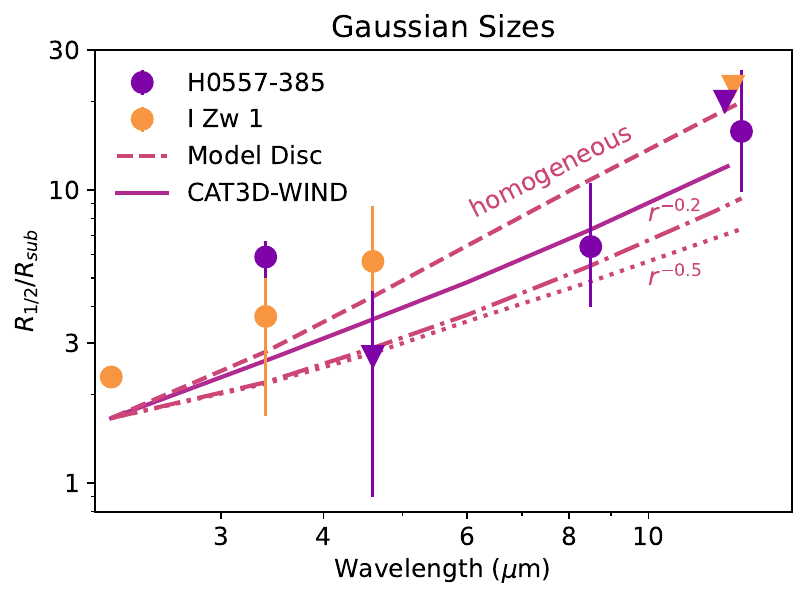}
    \caption{The scaled Gaussian half light radii shown as a function of wavelength, with upside down triangles marking upper limits. Errors are derived solely from the interferometric sizes, uncertainties in $R_{\mathrm{sub}}$ and $L_\mathrm{bol}$ have not been considered. Also shown are the Gaussian half light radii tracks for a disk with a radial dust distribution. The dashed line shows the size profile extracted from a homogeneous disk model, the dashed-dotted line a disk with a radial power law of $r^{-0.2}$, and the dotted line a power law with $r^{-0.5}$. The solid line shows the sizes extracted from a CAT3D-WIND model with a wind opening angle of $60^{\circ}$.}
    \label{fig:sizes}
\end{figure}

\begin{figure}
    \centering
    \includegraphics[width = 0.49\textwidth]{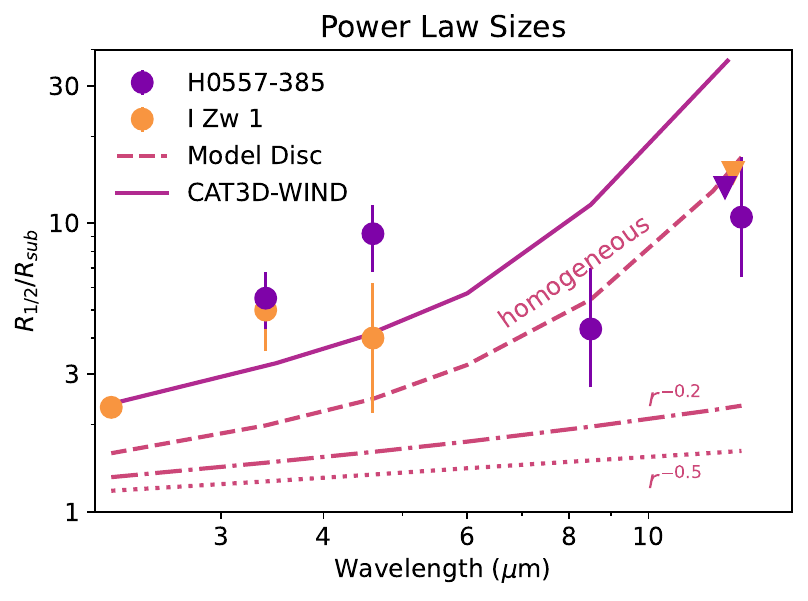}
    \caption{The scaled power law half light radii shown as a function of wavelength, with upside down triangles marking upper limits. The $K$-band half light radius is the derived from a single Gaussian fit, as discussed in Section \ref{data_grav}. Errors are derived solely from the interferometric sizes, uncertainties in $R_{\mathrm{sub}}$ and $L_\mathrm{bol}$ have not been considered. Also shown are the power law half light radii tracks for a disk with a radial dust distribution. The dashed line shows the size profile extracted from a homogeneous disk model, the dashed-dotted line a disk with a radial power law of $r^{-0.2}$, and the dotted line a power law with $r^{-0.5}$. The solid line shows the sizes extracted from a CAT3D-WIND model with a wind opening angle of $60^{\circ}$.}
    \label{fig:sizes_pl}
\end{figure}

For $N$-band sizes we used the results from \cite{burtscher2013}. Instead of a single Gaussian, the authors used a Gaussian and an unresolved point source to fit the MIDI correlated fluxes. Consequently, for source point fractions above 0.5, only an upper limit -- the maximum size of the unresolved region -- can be determined. This is the case for both of our objects. An alternative analysis of the MIDI data for H0557-385 is presented in \cite{kishimoto2011b}. Here, sizes were determined by fitting a power law instead of a Gaussian (see Section \ref{data_matisse}). As the authors illustrate in their Figure 5, depending on the size and spatial frequency, the HWHM of a Gaussian and $R_{1/2}$ of a power law show significant differences. Based on the visibilities, we converted the power law half light radii into Gaussian HWHM, using this figure. We multiplied the power law derived sizes by a factor of 1.5. At low spatial resolutions, power law and Gaussian + unresolved point source models lead to approximately the same results. Accordingly, we applied the same correction method to sizes from \cite{burtscher2013}, for which we determined the correction factor to be 1.5 as well. Uncorrected sizes from \cite{burtscher2013} and \cite{kishimoto2011b} are included in Table \ref{tab:sizes_pl} along our other power law measurements. Sizes corrected for a single Gaussian model are presented in Table \ref{tab:sizes_gauss}. In addition, Gaussian normalised half light radii as a function of wavelength are plotted in Figure \ref{fig:sizes}, and normalised power law half light radii are plotted in Figure \ref{fig:sizes_pl}. The $K$-band data point in the plot depicting the power law sizes (Figure \ref{fig:sizes_pl}) is the single Gaussian half light radius, as discussed in Section \ref{data_grav}.

Angular sizes and physical radii presented in this paper can be compared with results reported previously for other AGN. The (uncorrected for accretion disk contribution) angular FWHM of I Zw 1 in the $K$-band, 0.65 mas, corresponds to the value measured by \cite{gravity2024}. After correcting the physical radius for a thin ring geometry, as applied by the authors, our measurements also arrive at the same value at 0.59 pc. In the $LM$-band, sizes have been reported only for 2 other sources so far, none of which are Type 1s. Sizes have only been published for the nearest and brightest AGNs, NGC 1068 and Circinus, both of which are Type 2s. In NGC 1068, the $LM$-band is measured to have a size of $1.7 \times 0.9\:\mathrm{pc}$, which, assuming a sublimation radius of 0.15 pc (in line with our analysis), corresponds to $\sim 11.3 \times 6\:R_{\mathrm{sub}}$ \citep{burtscher2013,gamezrosas2022}. For Circinus, the analysis of the $LM$-band emission is more detailed. With an $R_{\mathrm{sub}} \sim 0.02\:\mathrm{pc}$, the $L$-band size is $\sim 6\times 3\:R_{\mathrm{sub}}$ ($0.12 \times 0.06\:\mathrm{pc}$) \citep{burtscher2013,isbell2023}. The $M$-band is modelled with two components: a large and a small one. The large component has an estimated size of $30\times 4.3\:R_{\mathrm{sub}}$ ($0.6 \times 0.08\:\mathrm{pc}$). Interestingly, the small component appears to be smaller than the $L$-band, just $5\times 2.2\:R_{\mathrm{sub}}$ ($0.1 \times 0.044\:\mathrm{pc}$). While these are on the same scales we are measuring, comparing the components, our results show sizes towards the larger end as measured by a power law, and towards the smaller end when measured by a single Gaussian model. It should be noted again, that our objects are type 1 AGN, meaning that more of the central hot dust emission is exposed to the observer. As such, smaller sizes compared to type 2s can be expected.

To put our results into the context of a model for the circumnuclear environment in these sources, we assume the dust distribution to be primarily located within or projected onto a disk, given the lack of significant polar extension. If a wind is present it must emerge from close to the disk with a wide opening angle or fully projected onto the disk. To test this, we adopt a simplified torus/disk model consisting of a power law dust distribution $\eta(r) \propto (r/r_\mathrm{in})^{-a}$ and a black body temperature distribution $T(r) = T_\mathrm{in} \cdot (r/r_\mathrm{in})^{1/2}$ with $T_\mathrm{in} = T(r_\mathrm{in}) = 1500\,\mathrm{K}$ \citep[e.g.][]{honig2011}. We calculated normalised interferometric sizes ($R_{1/2}/R_{\mathrm{sub}}$) as a function of wavelength for $a=0$ (homogeneous dust distribution), $a=0.2$ and $a=0.5$. Sizes from the model images have been extracted in the same way as for the observations (see Section~\ref{data_matisse}). For the 1D Gaussian model, we assume that $R_{\mathrm{sub}}=0.6\cdot R_K$ where $R_K$ is the observed interferometric $K$-band size. This accounts for the well-documented difference between the brightness-weighted interferometric size estimates compared to the ``true'' inner radii as determined from response-weighted reverberation mapping \citep[e.g.][]{kishimoto2007,kishimoto2011a,gravity2020b}.
Figure \ref{fig:sizes} shows the Gaussian size tracks for these model disks and in Figure \ref{fig:sizes_pl}, we plot the corresponding power law size tracks.

For a more realistic comparison, we also extracted the interferometric size as a function of wavelength from CAT3D-WIND images of a disk and a wind with a large opening angle. We used a wind opening angle of $60^{\circ}$ with all other parameter values taken from the CAT3D-WIND best fit model at $30^{\circ}$ inclination (Table \ref{tab:mod_res}). The same method as for the simple disks was used to calculate the interferometric sizes here. Interferometric sizes of the CAT3D-WIND model calculated using a 1D Gaussian model are plotted as a solid line in Figure \ref{fig:sizes}, and those calculated using a power law model are plotted as a solid line in Figure \ref{fig:sizes_pl}.

\section{Discussion}\label{discussion}

\begin{figure*}
    \centering
    \includegraphics[width = \textwidth]{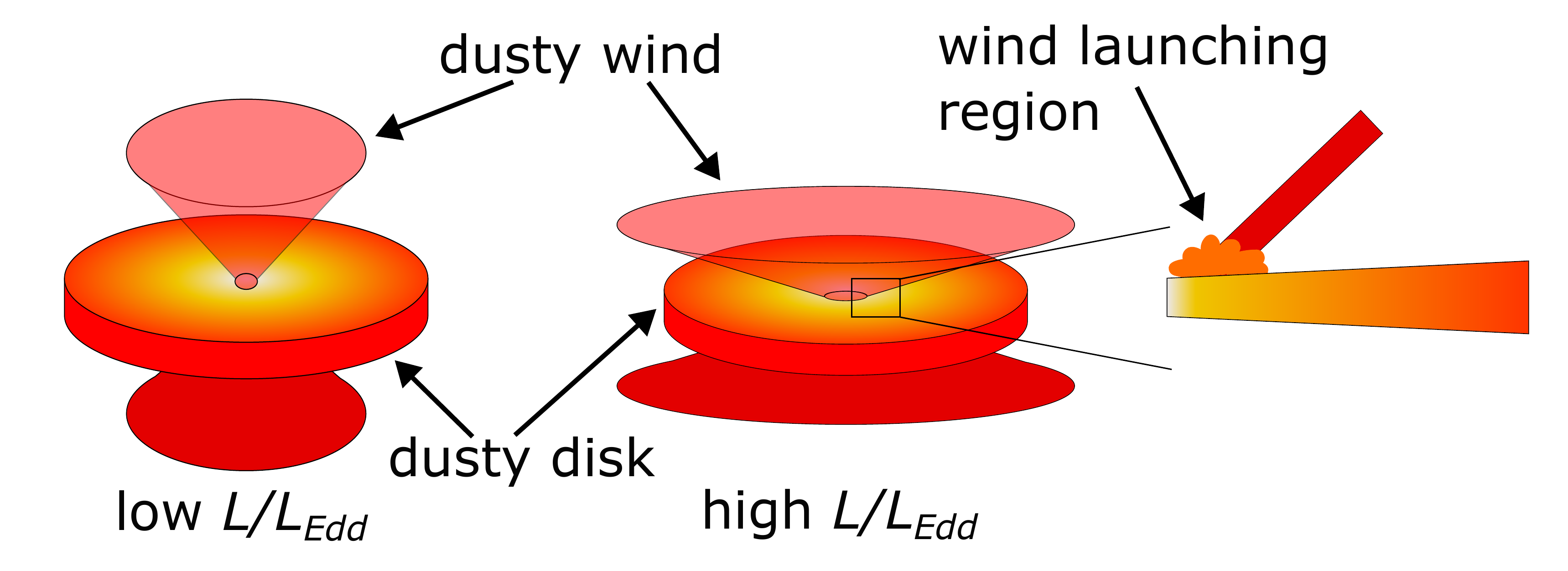}
    \caption{This illustration is an approximation of the torus structure in low (left) and high Eddington ratio AGNs (right), as the opening angle of the wind outflow cone increases. It also depicts a vertical slice of disk and wind at the inner boundary of the disk. Here, infrared radiation pressure launches the wind off of the disk. The colour gradient from white/yellow to red depicts the temperature gradient of the dust. Credit: Rowan Dayton-Oxland}
    \label{fig:diagram}
\end{figure*}

In this section, we test and discuss the disk+wind model for the dusty torus as described in \citet{honig2019}. Particularly, we look at the consequence of high Eddington ratios on this model: as the Eddington ratio increases, theory predicts that the opening angle of the dusty wind will increase, changing it from a \textit{polar} to a predominantly \textit{equatorial} direction. This model is illustrated in Figure \ref{fig:diagram}.


\subsection{No polar wind}

Our modelling results do not indicate the presence of a prominent polar structure in H0557-835. In Figure \ref{fig:model_obs}, the measured visibilities are compared to those for the CAT3D and CAT3D-WIND models best fitting the SED. The model visibilities were extracted from the corresponding radiative transfer model images at $3.4\:\umu\mathrm{m}$ and $4.6\:\umu\mathrm{m}$. The best fit parameters show that while the CAT3D iteration does not have a wind by construction, neither does the CAT3D-WIND iteration have a significant wind. Here, the wind contains less than a third of the entire infrared-emitting dust and has the largest opening angle available in the model, of 45 degrees. The shaded regions in the plots encompass the entire range of visibilities covered in perpendicular directions when simulating the interferometric observations of the images. They are narrow for both iterations, showing very little directional dependence, which is also reflected in our observations. In addition, the model results with and without wind are very similar in the visibility space, indicating that the addition of a weak polar wind does not make a significant impact on interferometric observations. Even with more precise data, it would not be possible to distinguish between these two instances of models. There, we assume for further discussion that H0557-385 does not harbour any significant \textit{polar} wind.

\subsection{Inferring the dust distribution within $10\,R_\mathrm{sub}$}

Figure \ref{fig:sizes} shows the single Gaussian interferometric sizes of our two objects as a function of wavelength. It also shows the Gaussian half light radii for a dusty disk with a homogeneous radial dust distribution, and two radial power law dust distributions, $r^{-0.2}$ and $r^{-0.5}$, as well as those of a CAT3D-WIND model with a wind opening angle of $60^{\circ}$ (as described at the end of Section \ref{int_size}). The CAT3D-WIND model lies between the homogeneous disk and the disk with $r^{-0.2}$ profile. H0557-385, in purple, has a large $L$-band size, clearly above the homogeneous dusty disk at that point. As wavelength increases, the relative sizes appear to flatten out (or even possibly dip in the $M$-band), within $\sim 10\,R_{\mathrm{sub}}$. The size then increases almost up to the homogeneous disk track again at the longest wavelengths of $12 - 13\:\umu\mathrm{m}$. The sizes of I Zw 1 in orange consistently follow the profile of the homogeneous disk, while staying just above it. The $KLM$-bands are concentrated within sizes below $\sim 5\,R_{\mathrm{sub}}$.

The power law interferometric sizes as a function of wavelength are presented in Figure \ref{fig:sizes_pl}. Power law half light radii for a dusty disk with a homogeneous radial dust distribution, a radial power law dust distribution with $r^{-0.2}$, and a radial power last dust distribution with $r^{-0.5}$ are also shown. The power law sizes of the CAT3D-WIND model with the large wind opening angle are consistently above those of the homogeneous disk. H0557-385 $LM$-band sizes (purple circles) are offset to larger sizes from the homogeneous disk distribution. At longer wavelengths, the sizes follow the homogeneous disk model more closely. Interestingly the $LM$-band sizes are consistent with the longer wavelength sizes implying that the bulk of the brightness in the 3.4$\:\mu\mathrm{m}$ to 12.5$\:\mu\mathrm{m}$ wavelength region emerges from the same spatial region ($\sim4-10\,R_{\mathrm{sub}}$). For I Zw 1 (orange circles), the short wavelength sizes in the $KLM$-bands roughly follow the CAT3D-WIND tracks, with sizes $\lesssim 5\,R_{\mathrm{sub}}$. Those tracks are slightly offset from the homogeneous disk but follow a similar slope. On the other other hand, the upper limit in the $N$-band indicates a deviation from the shorter wavelength trend towards a flatter dust distribution (approximately parallel to the $r^{-0.2}$ track).

It is worth highlighting that both the Gaussian model sizes as well as those extracted from the more physically motivated power-law model are consistent with each other within errors. This demonstrates that the analysis of our results is not dependent on the chosen model. Both approaches imply that a significant fraction of the total infrared emission is concentrated at about the radius constrained by the $LM$-bands.

\subsection{Where is the dust wind -- evidence for a wind launching region}

The presence of a dusty wind is often inferred through a \emph{polar} elongation in mid-IR interferometry and imaging \citep[e.g.][]{honig2012,asmus2016}. Both I Zw 1 and H0557-385 show no evidence of polar elongations in the $N$-band \citep{burtscher2013,lopez-gonzaga2016}, although we expect the wind emission to dominate the mid-IR \citep{tristram2014,isbell2022}. Radiative hydrodynamic (RHD) simulations of the dust around AGN show that due to anisotropy of the accretion disk radiation an increase in the accretion will increase the opening angle of the polar dust cone \citep[e.g.][]{williamson2020}. In this picture, higher accretion rates will flatten out the winds into the equatorial direction close to the disk (Figure \ref{fig:diagram}). We can now test if this model is consistent with our observations.

In H0557-385, the sizes 3.4 and $12.5\:\mu\mathrm{m}$ are more or less independent of wavelength. This implies that the bulk of the emission is concentrated within the $\sim 3-10\,R_{\mathrm{sub}}$ region. In I Zw 1, this flattening is prominently seen in the power law sizes, constrained to within $\sim 5\,R_{\mathrm{sub}}$ in the $KLM$ bands. Within the model of a radiatively driven wind, these observations are consistent with a wind launching region. In this region, a combination of infrared and AGN radiation pressure blows dust off of the disk into the wind (Figure \ref{fig:diagram}). This region represents a conglomeration of the dust in the disk (``puffed-up disk'') and would dominate the emission in the bands corresponding to the size of this region. Both H0557-385 and I Zw 1 show such a flattening in sizes at the expected distances \citep{honig2017}.

Puffed-up regions are regularly seen in young stellar objects \citep{dullemond2001,honig2019}. This bulge in the dust distribution will throw a shadow on the dust in the disk behind it, decreasing the temperature of that dust. As a result, sizes measured at longer wavelengths will be artificially decreased. This will lead to a ``bump feature'' in the size-wavelength relation where sizes remain constant.


\subsection{Where is the dust wind -- equatorial outflow or blow out?}

As mentioned previously, one important factor in the dusty wind model is the dependence of the wind direction on Eddington ratio. Both I Zw 1 and H0557-385 are high-Eddington ratio sources. As such, the model would predict an equatorial wind. This may appear as if the polar wind seen in lower Eddington ratio sources is blown out. Evidence for such blow out has been found in mid-infrared and sub-mm observations of some local AGN as shown in \citet{alonso-herrero2021} and \citet{garcia-burrilo2021}. Indeed, the objects that appear to lack a polar wind preferentially have either higher luminosities and/or accretion rates than typical Seyfert galaxies or are very low luminosity AGN where significant radiative feedback is not expected.

For our data, aside from the absence of a polar wind, we find that CAT3D-WIND models with large wind opening angles match the H0557-385 observations relatively well, in particular for our Gaussian size estimates. They do not reproduce the puffed-up region as it is not included in the model. While not being proof of an equatorial wind, this highlights that our observations are consistent with such a scenario. Interestingly, when focusing on the Gaussian sizes, I Zw 1 at higher Eddington rate follows closer the homogeneous disk model than H0557-385. This may be interpreted as a sign of a stronger blow-out of the wind or flattening of the wind launching region, in line with the overall radiatively driven dusty wind picture.

\subsection{The NIR dust structure in I Zw 1}

In addition, we want to discuss the $K$-band size of I Zw 1 and its implications for the NIR dust structure. Previous work has shown that the NIR emission region, the hot dust, is likely in the shape of a thin ring \citep{kishimoto2011a,gravity2020b,gravity2024}. For consistency with previous works, we also assume here that this is the case. To correct for this, we divide our Gaussian size by a factor $\sqrt{\ln{2}}$. The thin ring size is then $2.8 \,R_{\mathrm{sub}}$. The normalised size of the hot dust ($R/R_{\mathrm{sub}}$) can be used to get a qualitative picture of the dust emissivity and density slopes \citep{kishimoto2011b, gravity2020b}. If more hot dust is concentrated close to the sublimation radius, then $R/R_{\mathrm{sub}} \rightarrow 1$ -- the dust emissivity and density slopes are steep. On the other hand, a more spread out NIR emitting region will result in a large value of $R/R_{\mathrm{sub}}$. Given that the $K$-band value of $R/R_{\mathrm{sub}}$ in I Zw 1 is relatively large, this is in agreement with a shallow dust emissivity and density slope, and an extended NIR emitting region. This is in contrast to results found by \cite{kishimoto2011a} and \cite{gravity2020b}, who looked at the dependency of $R/R_{\mathrm{sub}}$ on the bolometric luminosity. They found that in general, $R/R_{\mathrm{sub}}$ decreases with increasing luminosity, meaning that emissivity and density slopes get steeper with higher luminosity. However, I Zw 1 is in the first quartile of normalised sizes when put in context of the \cite{gravity2020b} sample. In its corresponding luminosity bracket, it is larger than any other object. This suggest that there might be secondary effects driving the NIR dust structure, such as the Eddington ratio. 


Another explanation for the comparatively large relative $K$-band size in I Zw 1 is the uncertainty in the sublimation radius. While we adopted $R_{\mathrm{sub}} = 0.18\:\mathrm{pc}$, the 1 dex spread in $L_{\mathrm{bol}}$ allows for an up to 50\% increase in the size of the sublimation radius. With a sublimation radius of $\sim 0.3\:\mathrm{pc}$, the relative $K$-band size will decrease to $\sim 1.5\,R_{\mathrm{sub}}$, much more similar to other objects. However, going to the other extreme of $L_{\mathrm{bol}}$ decreases the sublimation radius even further, resulting in a relative $K$-band size of $\sim 4\,R_{\mathrm{sub}}$. Uncertainties in the calculation of the bolometric luminosity and the sublimation radius can have large effects on the interpretation of the results.


\section{Conclusions}\label{conclusions}
In this paper, we present the fist VLTI/MATISSE data of Type 1 AGN, focusing on the infrared structure of highly accreting sources. 
We used Gaussian and power law models to fit the interferometric data to recover the sizes from our MATISSE $LM$-band and archival GRAVITY $K$-band data. Using these data together with prior results from longer wavelength MIDI data, we were able to construct a multi-wavelength view of the dust structure. This shows that:
\begin{enumerate}
    \item There is no evidence for a polar wind in both objects. This could be either due to a wind being launched equatorially, projection effects, or the wind region being blown out at high Eddington ratios.
    \item Interferometric data implies a preferentially disky equatorial dust distribution in both objects.
    \item We find evidence that the near- to mid-IR emission is concentrated in the disk plane at distances $\sim3-10\,R_\mathrm{sub}$, which we interpret as signs of a puffed-up wind launching region.
\end{enumerate}
The presence of a puffed up inner region with absence of a \textit{polar} wind in these high Eddington ratio objects is notably different to previously observed lower accreting Seyferts. However, it matches predictions of a radiation pressure driven wind model. To further examine the relationship between the behaviour of the dusty ‘torus’ and the Eddington ratio, we need to expand our current sample of multi-wavelength interferometrically observed high-Eddington objects. The use of GRA4MAT in mid-IR observations will increase the quality of data and enable new $N$-band observations which will provide us with phases for the first time. Currently ongoing upgrades of GRAVITY into GRAVITY+ will further increase capabilities.

\section*{Acknowledgements}
We thank the anonymous referee for their helpful and constructive comments that improved the manuscript. FD acknowledges support from the Science and Technology Facilities Council PhD Studentship and the ESO Studentship Program. JL acknowledges partial financial support through CEFIPRA program 6504-2. This work was supported by the French government through the National Research Agency (ANR) with funding grant ANR AGN\_MELBa (ANR-21-CE31-0011). SFH acknowledges support through STFC grants ST/V001000/1 and ST/Y001656/1. MK acknowledges partial support by JSPS grants 20K04029/24K00679. We thank the GRAVITY AGN team for sharing their $K$ band interferometry data on I Zw 1 ahead of their own publication.

This work made use of Astropy a community-developed core Python package and an ecosystem of tools and resources for astronomy \citep{astropy2013, astropy2018, astropy2022}. 

\section*{Data Availability}

This work is based on observations collected at the European Southern Observatory under ESO programs 085.C-0172, 0105.B-0346(A), and 1103.D-0626(C). The raw data is publicly available from the ESO Science Archive Facility at \href{http://archive.eso.org/cms.html}{archive.eso.org}. Reduced MATISSE visibilities are available from FD upon request. 



\bibliographystyle{mnras}
\bibliography{references}



\appendix

\section{MATISSE Data Reduction}\label{matisse data reduction}
\begin{figure*}
    \centering
    \includegraphics[width=0.49\textwidth]{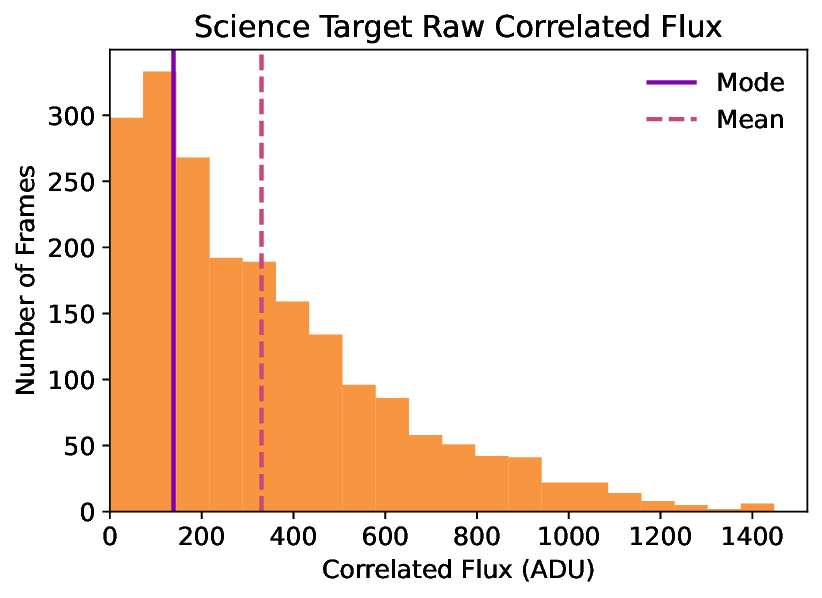}%
    \includegraphics[width=0.49\textwidth]{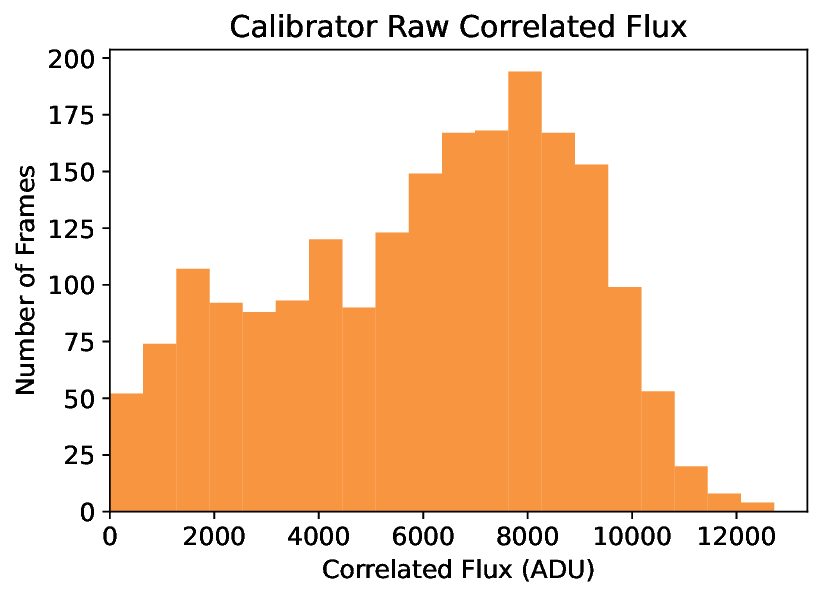}
    \caption{The raw correlated flux distributions of both the science target (left) and the red calibrator (right). As the science target is faint, its distribution is heavily skewed towards lower fluxes. We have also indicated the position of the mode and the mean of the science target distribution. Here, the mean overestimates the mode by over two times. }
    \label{fig:raw_dist}
\end{figure*}

\begin{figure*}
    \centering
    \includegraphics[width=0.49\textwidth]{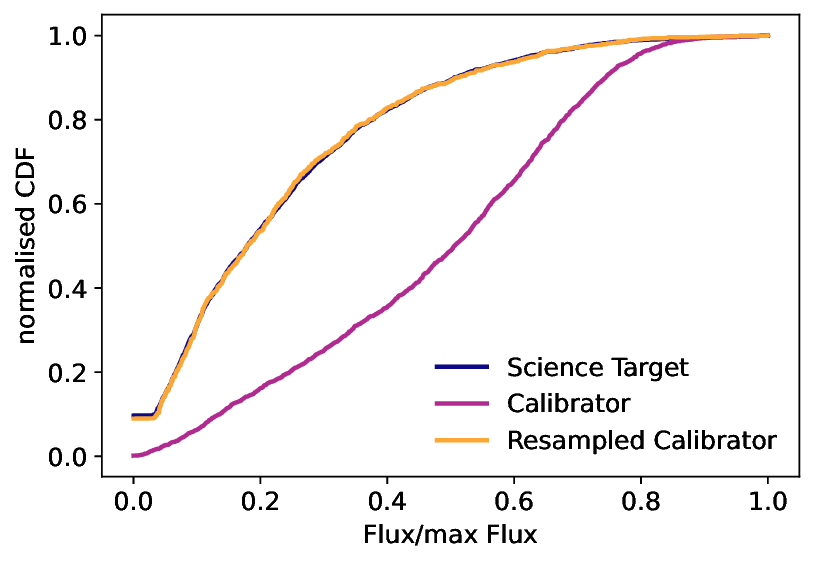}%
    \includegraphics[width=0.49\textwidth]{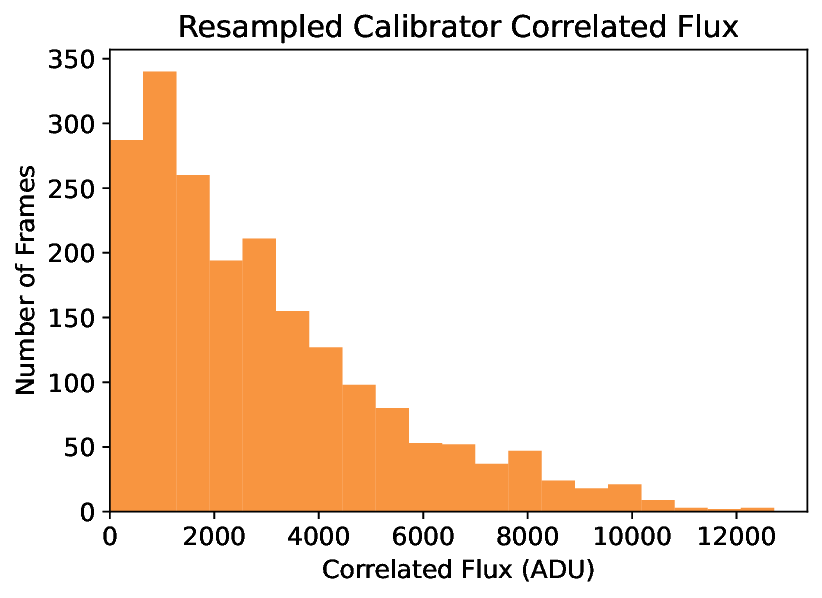}
    \caption{Plots illustrating our process of resampling the calibrator distribution to match that of the science target, to remove instrumental effects. On the left we have plotted the cumulative distribution functions (CDF): frames from the original calibrator distribution (magenta line) are chosen such that the CDF matches the science target (dark blue line), resulting in the resampled calibrator distribution (yellow line). The resampled calibrator distribution is also plotted on the right, and is almost equivalent to the science target in Figure \ref{fig:raw_dist}.}
    \label{fig:resampling}
\end{figure*}

\begin{figure*}
    \centering
    \includegraphics[width = 0.48\textwidth]{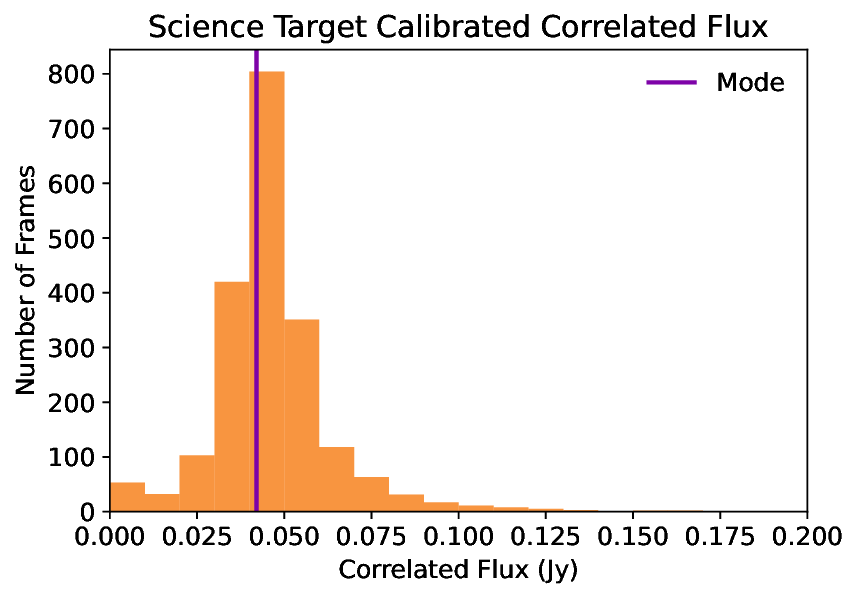}
    \caption{The calibrated correlated flux of the science target after deconvolving the science target with the red calibrator, which produces a narrowly defined distribution, well characterised by the mode.}
    \label{fig:targ_cal}
\end{figure*}

\begin{figure*}
    \centering
    \includegraphics[width=0.49\textwidth]{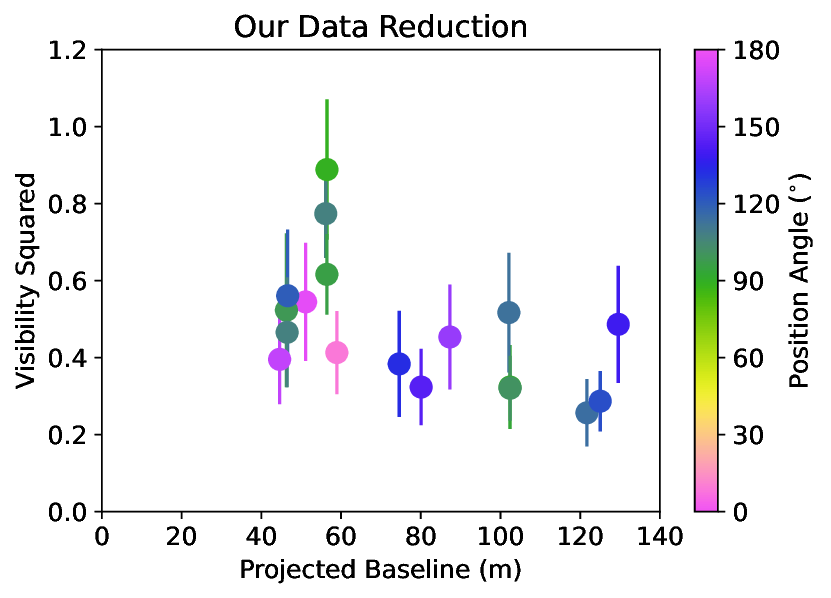}%
    \includegraphics[width=0.49\textwidth]{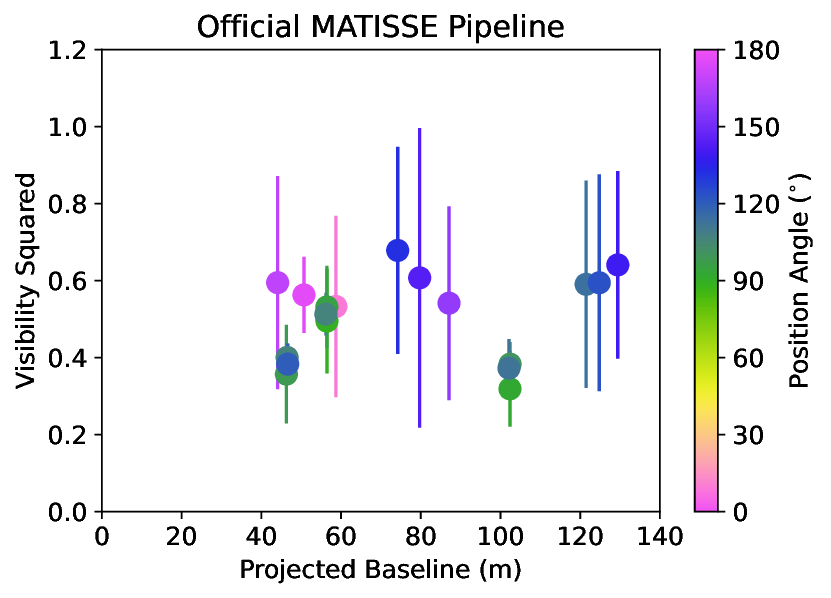}
    \caption{Plots illustrating the outputs of our data reduction (left) and the output of the official MATISSE pipeline (right, v1.7.6). While our data reduction certainly increases the scatter, it significantly decreases the errors. In addition, scatter in the pipeline is only small because of the removal of outliers and using the mean. Our reduction shows the source to be partially resolved while the pipeline portrays ambiguous results.}
    \label{fig:comparison}
\end{figure*}

We will start by defining the problem we face in the reduction of interferometric data of ‘faint’ AGNs. Firstly, we have the fundamental challenge that the correlated flux (and therefore the visibility) is positively biased since we cannot measure negative fluxes. To compound this problem, these objects also have a correlated flux distribution skewed towards lower fluxes, with a long high flux tail, due to the poor AO performance resulting in lower coherence. Figure \ref{fig:raw_dist} clearly illustrates that the average (as calculated by the MATISSE pipeline) of this distribution will additionally overestimate the correlated flux and the visibility. Similarly, while brighter, the red calibrators we use are also comparatively low flux. More importantly, their flux distribution is drawn from a different parent distribution as that of the science target, preventing direct comparison (see Figure \ref{fig:raw_dist}). Instrumental and AO performance strongly affects this. A peculiarity of MATISSE is the presence of the BCD (beam commuting device) which introduces additional offsets depending on the setting. To account for this effect, observations taken with different BCD positions must be reduced and calibrated independently. However, with the determination of the BCD corrections, we are able to lever the power of large samples and combine different observations to reduce noise. In addition, we also masked out lower signal-to-noise wavelengths bins on a per baseline per frame basis to reduce the amount of unnecessarily discarded data.

Crucially, we remove instrumental effects by deconvolving the science target with the red calibrator. Practically, we resampled the calibrator distribution to look like the target distribution. In effect this is represented mathematically as
\begin{equation}
    \mathrm{TD} = \mathrm{CD} \circledast \mathrm{TF} \Rightarrow \mathrm{TD}(p) = \int_{D}\mathrm{CD}(p')\cdot\mathrm{TF}(p-p') dp'
\end{equation}
where the calibrator distribution CD is convolved with the transfer function TF to give the target distribution TD (that of the science object). This is accomplished by shifting the calibrator’s cumulative distribution function, as shown in Figure \ref{fig:resampling}. This effectively matches the red calibrator’s distribution to that of the science target’s distribution, resulting in two well defined correlated flux distributions, which we calibrate frame by frame. Since the flux of the calibrator is higher, we match the frames between the distributions based on their normalised distance to the mode. The $L$- and $M$-bands are exposed in the same frame, so instrumental effects will be the same. Accordingly, resampling of the distribution is solely done using the $L$-band as it has a higher signal-to-noise ratio. Frames are matched and calibrated separately for the $M$-band. The calibrated correlated flux value for each baseline is then found by taking the mode of the calibrated flux distribution (see Figure \ref{fig:targ_cal}). Errors are estimated using the width of the distribution. The outcome of using this technique can be seen in Figure \ref{fig:comparison}, our technique clearly showing a partially resolved source while the pipeline does not. This method enables us to use all of the available data and is a statistically robust technique to reduce interferometric data of faint AGNs.

\section{SED Data}

\begin{table*}
    \centering
    \begin{tabular}{c|cc|cccc}
         Type & Wavelength ($\umu\mathrm{m}$) & Flux ($10^{-14}\:\mathrm{W/m^2}$) & Instrument & Observation Date & Extraction Aperture Size & Reference\\
         \hline
         Photometry & 1.2 & $6.15\pm0.24$ & 2MASS & 2000-11-30 & $7"$ & \cite{shangguan2018} \\
         '' & 1.6 & $6.50\pm0.26$ & '' & '' & '' & '' \\
         '' & 2.2 & $7.94\pm0.16$ & '' & '' & '' & '' \\
         '' & 3.4 & $7.64\pm0.004$ & WISE & 2010-07-11 & $8".25$ &  '' \\
         '' & 4.6 & $8.62\pm0.005$ & '' & '' & '' & ''\\
         '' & 22.1 & $12.9\pm0.04$ & '' & 2010-07-10 & $16".5$ & '' \\
         '' & 70.0 & $9.59\pm0.03$ & Herschel/PACS & 2011-07-24 & $12"$ & '' \\
         Photometry & 8.99 & $10.1\pm0.91$ & VLT/VISIR & 2010-10-20 & $0.3"$ & \cite{asmus2014} \\
         '' & 11.74 & $12.2\pm0.57$ & Subaru/COMICS & 2006-10-04 & '' & '' \\
         '' & 11.88 & $11.3\pm1.3$ & VLT/VISIR & 2010-10-17 & '' & '' \\
         '' & 12.0 & $10.6\pm1.9$ & '' & -- & '' & '' \\
         Spectrum & $2.5 - 5.0$ & -- & AKARI/IRC & 2008-07-09 & $7.3"\times 1'$ & \cite{kim2015} \\
         Spectrum & $7.8 - 13.2$ & -- & VLT/VISIR & 2010-10-20 & $0.4"$ &  \cite{jensen2017}\\
         Spectrum & $5.5-35.0$ & -- & SPITZER/IRS & 2004-01-07 & $11"\times57"$& \cite{shi2014} \\
         Total Flux & 3.4 & $8.39\pm1.23$ & VLTI/MATISSE & 2021-09-25 & -- & This paper\\
         Correlated Flux & 3.4 & $6.67\pm0.75$ & '' & '' & -- & '' \\
         Total Flux & 4.6 & $11.5\pm3.02$ & '' & '' & -- & ''\\
         Correlated Flux & 4.6 & $10.3\pm1.66$ & '' & '' & -- & '' \\
         Correlated Flux & 9.0 & $7.57\pm1.07$ & VLTI/MIDI & 2010-08-25 & -- & \cite{burtscher2013} \\
         '' & 12.0 & $8.44\pm1.77$ & '' & '' & -- & '' \\
    \end{tabular}
    \caption{SED data for I Zw 1.}
    \label{tab:data_izw1}
\end{table*}

\begin{table*}
    \centering
    \begin{tabular}{c|cc|cccc}
         Type & Wavelength ($\umu\mathrm{m}$) & Flux ($10^{-14}\:\mathrm{W/m^2}$) & Instrument & Observation Date & Extraction Aperture Size & Reference\\
         \hline
         Photometry & 1.0 &$1.29\pm0.06$ & UKIRT/WFCAM & 2009-12-13 & $\ast$ & \cite{kishimoto2011b}\\
         '' & 1.2 & $1.73\pm0.09$ & ''& ''&''& ''\\
         '' & 1.6 & $2.79\pm0.14$ & ''& ''&''& ''\\
         '' & 2.2 & $4.99\pm0.25$ & ''& ''&''& ''\\
         Photometry & 3.6 & $11.7\pm0.06$ & SPITZER/IRAC & 2008-10-31 & $\ast$ & '' \\
         '' & 4.5 & $13.0\pm0.06$ & ''& ''&''& ''\\
         '' & 5.7 & $14.1\pm0.13$ & ''& ''&''& ''\\
         '' & 7.9 & $13.1\pm0.06$ & ''& ''&''& ''\\
         Photometry & 8.59 & $11.5\pm0.38$ & VLT/VISIR & 2009-09-07 & $0.33"$& \cite{asmus2014}\\
         '' & 10.49 & $9.78\pm0.54$ & '' & 2009-09-30 & ''& ''\\
         '' & 11.88 & $10.4\pm1.49$ & '' & 2009-09-07 & ''& ''\\
         '' & 11.88 & $9.65\pm0.42$ & '' & 2009-09-30 & ''& ''\\
         '' & 12.0 & $10.6\pm1.07$ & '' & -- & ''& ''\\
         '' & 12.81 & $11.2\pm1.34$ & '' & '' & ''& ''\\
         Photometry & 12.0 & $13.2\pm0.79$ & IRAS & 1983-03-27 & -- & \cite{moshir1990}\\
         '' & 22.0 & $9.33\pm0.47$ & ''& ''&''& ''\\
         '' & 52.0 & $1.86\pm0.20$ & ''& ''&''& ''\\
         Spectrum & $5.2 - 36.8$ & -- & SPITZER/IRS & 2007-10-06 &$11"\times57"$& Spitzer Heritage Archive \\
         Total Flux & 3.4 & $7.34\pm1.81$ & VLTI/MATISSE & 2021-09-25 & -- & This paper\\
         Correlated Flux & 3.4 & $4.86\pm1.03$ & '' & '' & -- & '' \\
         Total Flux & 4.6 & $12.5\pm4.59$ & '' & '' & -- & ''\\
         Correlated Flux & 4.6 & $8.07\pm1.51$ & '' & '' & -- & '' \\
         Correlated Flux & 9.0 & $3.92\pm0.97$ & VLTI/MIDI & 2009-08-01 & -- & \cite{burtscher2013} \\
         '' & 12.0 & $6.92\pm1.30$ & '' & '' & -- & '' \\
    \end{tabular}
    \caption{SED data for H0557-385. $\ast$: data was reduced to remove host galaxy contamination. For further details see \citet{kishimoto2007,kishimoto2011b}.}
    \label{tab:data_h0557-385}
\end{table*}

\section{SED Modelling Results}

\begin{table*}
    \centering
    \begin{tabular}{cccccccccccc}
        Model & Inclination ($^{\circ}$) & $R_{\mathrm{sub}}$ (pc) & $a$ & $N_0$ & $h$ & $a_w$ & $\theta_w$ ($^{\circ}$) & $\sigma_\theta$ ($^{\circ}$) & $f_{wd}$ & $\tau$ & $\chi^2_{red}$\\
        \hline
        CAT3D & 15 & 0.10 & -2.25 & 7.5 & 0.25 & -1.50 & 30 & 15 & 0.0 & 2.6 & 87\\
        CAT3D & 30 & 0.10 & -2.25 & 7.5 & 0.25 & -1.50 & 30 & 15 & 0.0 & 2.3 & 82\\
        CAT3D & 45 & 0.12 & -2.25 & 5.0 & 0.25 & -1.50 & 30 & 15 & 0.0 & 2.2 & 83\\
        CAT3D-WIND & 15 & 0.12 & -2.5 & 5.0 & 0.3 & -2.0 & 30 & 15 & 2.0 & 2.2 & 167\\
        CAT3D-WIND & 30 & 0.12 & -2.5 & 5.0 & 0.4 & -1.5 & 45 & 10 & 0.45 & 2.2 & 167\\
        CAT3D-WIND & 45 & 0.12 & -2.5 & 5.0 & 0.4 & -1.5 & 45 & 7 & 0.45 & 2.3 & 163\\
    \end{tabular}
    \caption{CAT3D and CAT3D-WIND SED modelling results for H0557-385. Model parameters are as follows: inclination of the AGN, sublimation radius $R_{\mathrm{sub}}$, index $a$ of the radial dust distribution power law in the disk, line-of-sight cloud number $N_0$, scale height $h$, index $a_w$ of the wind dust distribution power law, wind half-opening angle $\theta_w$, width of the wind cone $\sigma_\theta$, ratio of wind-to-disk dust clouds $f_{wd}$ \citep[for more information see ][]{honig2010,honig2017}. Additionally, we consider the host galaxy extinction with an optical depth of $\tau$. For CAT3D fits, $a_w$, $\theta_w$, and $\sigma_\theta$ were held constant with $f_{wd}=0$ to switch off the wind.}
    \label{tab:mod_res}
\end{table*}


\bsp	
\label{lastpage}
\end{document}